\documentclass[oldversion]{aa}
\usepackage{txfonts}
\usepackage{graphicx}
\usepackage[dvips]{color}
\usepackage{natbib}
\usepackage{supertabular}
\usepackage{color}

\bibpunct{(}{)}{;}{a}{}{,}

\begin{document}
\title{Results of optical monitoring of 5 SDSS double QSOs with the Nordic Optical Telescope}
\author{Danuta Paraficz \inst{1}, Jens Hjorth\inst{1}, \and \'Ard{\'\i}s El{\'\i}asd\'ottir \inst{1,2} }

\institute{Dark Cosmology Centre, Niels Bohr Institute, University of Copenhagen,
		Juliane Maries Vej 30, DK-2100 Copenhagen, Denmark
              \email{danutas@dark-cosmology.dk}
\and
Department of Astrophysical Sciences, Princeton University, Princeton, NJ08544, USA
}

\date{Received November 20, 2008}

\authorrunning{Danuta Paraficz} 
\titlerunning{Results of optical monitoring of 5 SDSS double QSOs with the Nordic Optical Telescope}

\abstract{We present optical R-band light curves of five SDSS double QSOs 
(SDSS J0903+5028, SDSS J1001+5027, SDSS J1206+4332, SDSS J1353+1138, 
SDSS J1335+0118) obtained from monitoring at the Nordic Optical Telescope (NOT) 
between September 2005 and September 2007. We also present analytical and 
pixelated modeling of the observed systems.
For SDSS J1206+4332, we measured the time delay to be 
$\Delta \tau = 116^{+ 4 }_{-5}$ days, which, for a Singular 
Isothermal Ellipsoid model, corresponds to a
Hubble constant of $73^{+3}_{-4}$ $\rm km\ s^{-1}\ Mpc^{-1}$.
Simultaneous pixeleted modeling of five other systems for which 
a time delay has now been previously measured at the NOT leads to 
$ H_0 =61.5^{+8}_{-4}$ $\rm km\ s^{-1}\ Mpc^{-1}$. Finally, by comparing lightcurves of the two images of each system,
suitably shifted by the predicted or observed time-delays, we 
found no evidence for microlensing variability over the course of
the monitoring period.
}


\maketitle

\section{Introduction}

A gravitationally lensed quasar is a quasar (QSO) which is lensed by a massive foreground object such as a galaxy or a group of galaxies.
If the lens is close enough to the line of sight then the quasar is strongly lensed and 
will be multiply imaged.
Due to the different travel times for each light path, any intrinsic variation of the quasar is observed in the images at different times.
This time difference, referred to as the time delay, can be measured by comparing the light curves of the images. As first shown by \citet{Refsdal:1964} the Hubble constant can be determined from the time delay provided that the mass distribution is known. Conversely, for a fixed Hubble constant, the mass distribution of the lensing galaxy can be constrained.

Determining the Hubble constant from the time delays between light variations in multiple images of gravitationally lensed QSOs is a classical application of lensing in cosmology, but despite 25 years of effort, results remain inconclusive. This is partly due to incomplete knowledge of the mass distribution along the light path to the QSO and partly due to the perturbing effects of microlensing on sparsely sampled light curves of the underlying long-term variations.

To address these issues,
we have conducted a monitoring program at the Nordic Optical Telescope (NOT)  
with the aim of obtaining densely sampled light curves of five lensing systems 
(SDSS J0903+5028, SDSS J1001+5027, SDSS J1206+4332, SDSS J1353+1138, 
SDSS J1335+0118).  
Dense sampling is required to quantify the effects of microlensing
\citep{Paczynski:1986, Schild:1996, Paraficz:2006} and to maximize chances of
determining a time delay from a relatively short monitoring campaign.

Estimating the Hubble constant using time delay measurements is strongly dependent on the underlying mass distribution and hence the choice of lens model \citep{Oguri:2007}.                                                                                                                                                                                                                                                                                                                                                                                                                                                                                                                                                                                                               
Two different approaches to modeling lenses are commonly used.
The first one, the non-parametric method \citep{Saha:1997},  generates a large number of models which perfectly fit the data, each of them gives a different time delay  which can be then averaged. 
For the second method, the analytical method\citep{Keeton:2001}, one assumes physical properties
of the mass distribution of the lens. 
Comparison of the two approaches gives
a useful indication on the systematic errors in e.g. the Hubble constant
determined this way.

Luckily, it seems that simple lens models are very good first approximations to the real mass distributions of lenses \citep{Koopmans:2006}.  Therefore we have chosen to use the singular isothermal potential to analytically model all the lenses we observe.

In this paper we present the results of a monitoring campaign at the NOT of 5 doubly lensed quasars. We have measured the time delay  of one of the lensing systems, SDSS J1206+4332, by analyzing light curves of the two quasar images obtained from six months of monitoring, demonstrating the feasibility of short-term
monitoring for time-delay measurement.  Based on our measurement of the  
time delay and 5 other time delay measurements previously obtained at the NOT we have estimated  the Hubble constant.

Section 2 describes the details of our monitoring campaign and section 3 
introduces the observed targets.
The photometric technique based on image deconvolution is described in 
section 4 and the light curves of all 5 targets are presented in section 5.
In section 6 the time delay of SDSS J1206+4332 is determined. In section 7 
we  perform a microlensing search in the 5 quasar light curves and estimate 
upper limits to the microlensing signal. In section 8  we perform 
analytical and pixeleted modeling. Section 9  is devoted to 
simultaneous modeling of 5 NOT-determined time delay systems with the aim
of a joint Hubble constant estimate. In section 10 we 
 discuss the results.

In the paper we use a flat $\Lambda$CDM Universe, $\Omega_{\rm m}=0.3$ and $\Omega_{\Lambda}=0.7$.

\section{Observation}

The observations were gathered from monitoring programs carried out in the 
periods September--March 2005/2006, October--March 2006/2007 and 
April--September 2007 at the Nordic Optical Telescope (NOT); a 2.5-m telescope 
located at  Roque de los Muchachos in La Palma, Spain. The advantage of this 
telescope is its fairly flexible scheduling, which made almost nightly 
monitoring possible. Our targets were observed every night under all three 
operative modes at  the NOT: observer, service and technical. However there were severe obstacles preventing frequent sampling: bad weather, Guaranteed Time programs at the NOT and sensitivity of the deconvolution software to imperfectness of the data.  

The detectors used in the monitoring were chosen in order to obtain the most frequent sampling and the best image quality. Thus, we used ALFOSC (The Andalucia Faint Object Spectrograph and Camera; pixel scale $0\farcs189$) whenever it  was mounted on the telescope and 
StanCam (the Stand-by CCD Camera, which is permanently mounted at the NOT;
pixel scale $0\farcs176$), otherwise. The seeing varied from $0\farcs4$ to $3\farcs0$ with $1\farcs0$ being the most frequent value.

We observed the objects in the R band, only.
The pilot phase of the program involved monitoring of  a  gravitationally lensed system SDSS~J0903+5028 at $z=3.6$ \citep{Johnston:2003} for 10 minutes every night. This system contains two quasar images (see Table \ref{phot}) separated by $2\farcs8$ and  aligned on opposite sides of the lensing galaxy at the redshift $z=0.388$ \citep{Johnston:2003}.

 After the first observing season we decided to observe three lensed systems alternately, SDSS~J1001+5027 and SDSS~J1206+4332 together in one night for 5 minutes each, and 
SDSS~J0903+5028 every second night for 10 minutes. The third observing season was again divided in two groups.
We continued observing SDSS~J1001+5027 and SDSS~J1206+4332 for 5 minutes each every second night, alternately with two new targets, SDSS J1335+0118 (5 min) and SDSS~J1353+1138 (3 min). The exposure times were chosen so that the signal-to-noise ratio
of the fainter component of each system would be above 10. Finding charts of all the targets are presented in Figures \ref{field1} and \ref{field2}.

One of the major challenges in monitoring gravitationally lensed quasars 
is the lack of  prior knowledge of time delay.
Thus, time-delay measurements require well-sampled light curves with accurate 
photometry over a period of time substantially longer than the time delay. 
Because the time delay is  not known, one has to calculate a theoretical time 
delay before planning the observations. 

Assuming that the theoretical prediction of the time delay for a given system 
is a good approximation (which is not always the case)
there is still a question as to whether the quasar will 
vary during the period of monitoring and what will be the timescale and 
amplitude of the variations.  
Quasar brightness might vary from a day up to years independently of the mass of the black hole \citep{Wold:2007}. We know that the rapid variability implies a light source at very small distance from the black hole \citep{Webb:2000}, while variations on long time-scales are related to morphological changes of jets on parsec scales or to accretion-disk instabilities \citep{Vries:2006}.
However, it is impossible to predict whether a quasar will vary or not within a given period and timescale, which is why the success of each quasar monitoring is uncertain.

Another difficulty with gravitationally lensed quasar observations is microlensing by stars in the lensing galaxy. Microlensing can change the results by about $0.5-1.0$ magnitudes and it is completely unpredictable \citep{Chang:1979}.

Thus, in order to get a time delay one has to monitor quasars with high 
sampling, so that events like microlensing can be extracted from the quasar 
variability  in further analysis. Our monitoring program was designed to 
minimize this problem.

Observation planning, monitoring supervision and image reduction 
was made by the first author during her stay at the NOT. Data reduction was
performed using  IRAF reduction utilities.

\section{Targets}
The individual targets observed are introduced below.
The astrometry and redshifts of the quasar images and the lenses
are summarized in Table  \ref{phot}.

\subsection{SDSS J0903+5028}

SDSS J0903+5028, a doubly lensed quasar system, was discovered from the Sloan Digital Sky Survey by \citet{Johnston:2003}. Using the ARC  3.5 meter telescope it was found that the system has two quasar images separated by 2\farcs8 with the lens, a red galaxy ($z=0.388$), in between. Spectroscopic follow up observation at the Keck II telescope proved that the two objects are the images of one quasar at $z=3.6$.  \citet{Johnston:2003} concluded that other galaxies in the vicinity of the lensing galaxy might be gravitationally bound with the lens, adding external shear to the lensed system.

 \begin{figure}
\centering
\resizebox{\hsize}{!}{\includegraphics{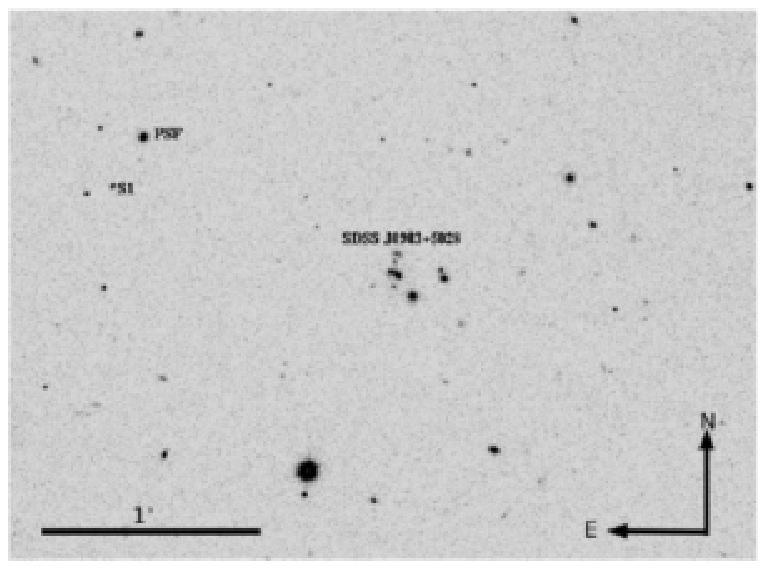}}
\resizebox{\hsize}{!}{\includegraphics{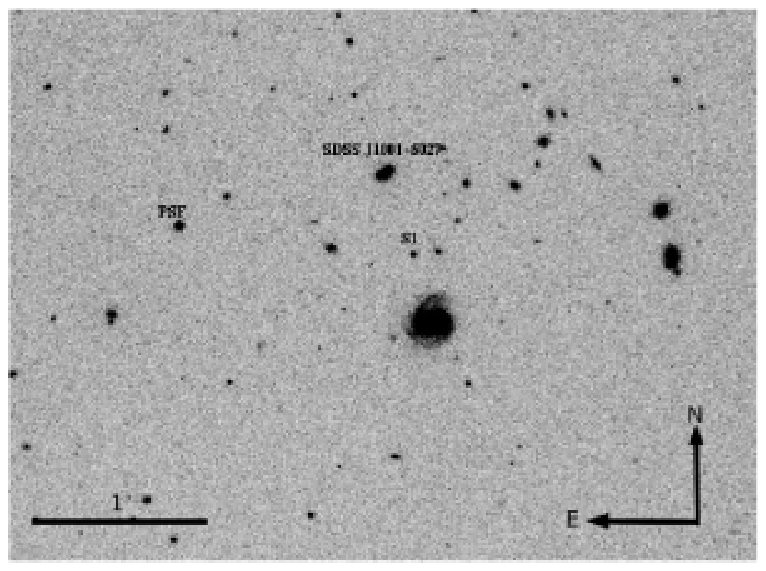}}
\resizebox{\hsize}{!}{\includegraphics{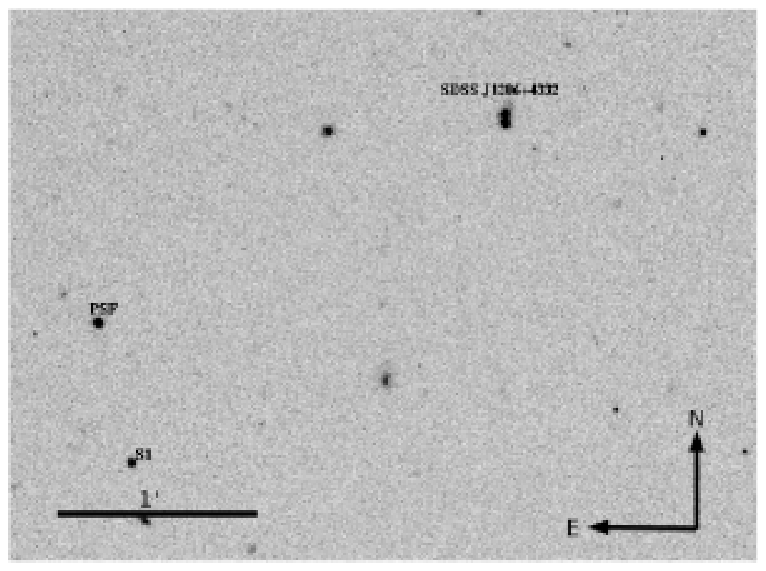}}
\caption{Finding chart of the three systems:  \textbf{Top}: SDSS J0903+5028, \textbf{Middle}: SDSS J1001+5027, \textbf{Bottom}: SDSS J1206+4332.  The reference star (S1) and the star used to model the PSF are indicated. }
\label{field1}
\end{figure}

\begin{table}[!ht]
\label{phot}
\centering
\begin{tabular}{  c c c c }
\hline\hline
&SDSS J0903+5028&\\
\hline

Object&R.A. (J2000.0) &Dec. (J2000.0) &$z$ \\
\hline
A&09 03 35.13& +50 28 20.21& 3.6 \\
B&09 03 34.88& +50 28 18.75& 3.6 \\
G&09 03 34.93& +50 28 19.53& 0.388\\
\hline\hline
&SDSS J1001+5027&\\
\hline
Object&R.A. (J2000.0) &Dec. (J2000.0) &$z$ \\
\hline
A&10 01 28.61& +50 27 56.9& 1.838  \\
B&10 01 28.35& +50 27 58.5& 1.838 \\
\hline
&x[arcsec]&y[arcsec]&$z$\\
\hline
G1&$1.779\pm0.049$&$0.857\pm0.123$& $0.2 \leq z\leq 0.5$\\
G2&$1.795\pm0.088$&$-0.700\pm0.053$&$0.2 \leq z\leq 0.5$\\
\hline\hline
&SDSS J1206+4332&\\
\hline
Object&R.A. (J2000.0) &Dec. (J2000.0) &$z$  \\
\hline
A&12 06 29.65&  +43 32 17.6& 1.789\\
B&12 06 29.65&  +43 32 20.6& 1.789 \\
\hline
&x[arcsec]&y[arcsec]&$z$ \\
\hline
G1&$-0.664\pm0.137$& $1.748\pm0.028$& 0.748\\
G2&$1.320\pm0.147$& $5.999\pm0.148$& $\geq0.7$\\
G3&$+2.052\pm0.200$& $2.397\pm0.152$& blue\\
\hline\hline
&SDSS J1335+0118&\\
\hline
&x[arcsec]&y[arcsec]&$z$\\
\hline
A&$0.000\pm0.001$&$0.000\pm0.001$& 1.57  \\
B&$-1.038\pm0.002$& $-1.165\pm0.002$& 1.57 \\
G&$-0.769\pm0.011$&$-0.757\pm0.011$& 0.44\\
\hline\hline
&SDSS J1353+1138&\\
\hline
&R.A. (J2000.0) &Decl. (J2000.0) &$z$\\
\hline
A&13 53 06.35& +11 38 04.81& 1.63  \\
B&13 53 06.08& +11 38 01.43& 1.63 \\
G&13 35 06.10& +11 38 00.39& $\sim0.3$\\
\hline
\end{tabular}
\caption{Astrometric properties and redshifts of the f\mbox{}ive lensed SDSS quasars: SDSS J0903+5028 from \citet{Johnston:2003}, SDSS J1001+5027 and SDSS J1206+4332 from \citet{Oguri:2005}, SDSS J1335+0118 from \citet{Oguri:2004} and SDSS J1353+1138 from \citet{Inada:2006}. Units of R.A. are hours, minutes and seconds, and units of Dec. are degrees, arcminutes, and arcseconds. All x,y positions are relative to the A component and their errors do not include the error of the image scale. The positive direction of the those coordinates are to the west and north, respectively.}

\end{table}

\subsection{SDSS J1001+5027 \& SDSS J1206+4332}
SDSS J1001+5027 and SDSS J1206+4332 were recognized as lensing candidates in a strong lensing survey  using the Sloan Digital Sky Survey. \citet{Oguri:2005} photometrically and spectroscopically conf\mbox{}irmed using the ARC 3.5 meter and University of Hawaii 2.2-meter telescopes,  that those two targets are indeed lensing systems.  They reported that SDSS J1001+5027 and SDSS J1206+4332 have image separation of 2\farcs86 and 2\farcs90 respectively with the source objects being quasars at $z=1.838$ and $z=1.789$.

\citet{Oguri:2005} noted that the two lensing galaxies of SDSS J1001+5027 have colors consistent with those of early-type galaxies at $0.2 \leq z\leq 0.5$. SDSS J1206+4332 appears to have three lensing galaxies where the main lens, G1, has an associated absorber at redshift $z=0.748$, the second galaxy, G2, is identifed as a high redshift galaxy $z \leq 0.7$ and, the third galaxy, G3, is a blue galaxy.  \citet{Oguri:2005} using the \textit{lensmodel} software \citep{Keeton:2001} modeled the two systems showing that they are both strongly affected by the potential of more than one galaxy.

\subsection{SDSS J1335+0118}

SDSS J1353+1138 is a doubly lensed system, discovered in the Sloan Digital Survey by \citet{Oguri:2004}.  The photometric follow up made at the Subaru 8.2-m and Keck I telescopes conf\mbox{}irmed that the system consists of two gravitationally lensed images separated by 1\farcs56 with a single lensing galaxy in the centre. Spectroscopic observations made at the ESO New Technology Telescope (NTT) showed that the A and B components of the system are images of a quasar at redshift $z=1.57$. \citet{Eigenbrod:2006b} reported that the lensing galaxy  is a low-redshift galaxy with $ z=0.44$.
                  
\begin{figure}
\centering
\resizebox{\hsize}{!}{\includegraphics{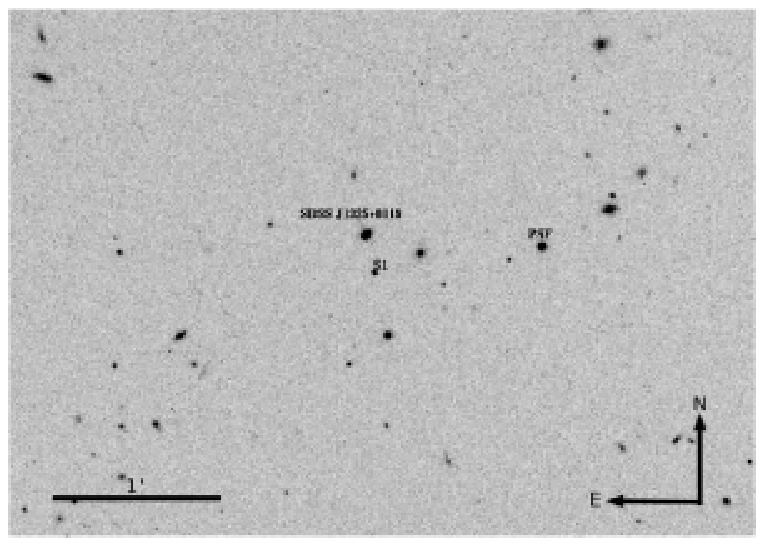}}
\resizebox{\hsize}{!}{\includegraphics{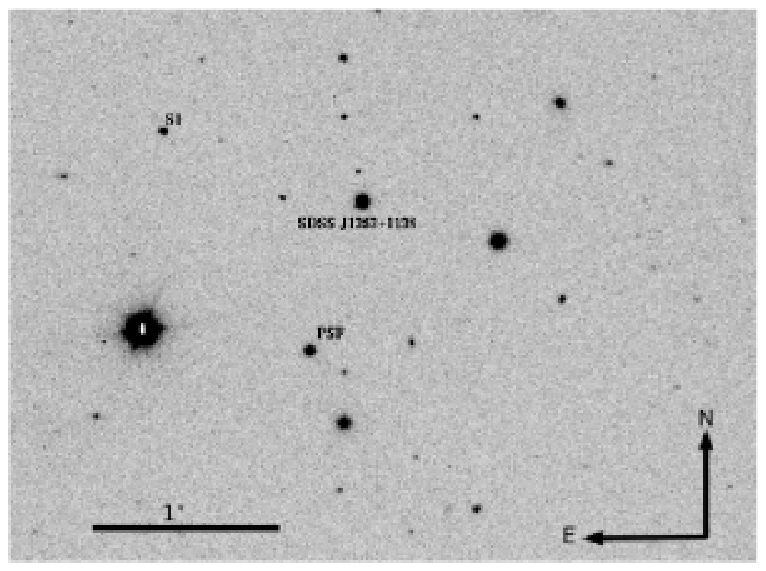}}
  \caption{Finding chart of the two systems: SDSS J1335+0118 and SDSS J1353+1138.  The reference star  and the star used to model the PSF are indicated}
    \label{field2}
\end{figure}

\subsection{SDSS J1353+1138}
\citet{Inada:2006} discovered the SDSS J1353+1138 lensed quasar system from SDSS lensed quasar survey. For imaging and spectroscopy \citet{Inada:2006} used the University of Hawaii 2.2-m telescope, the Keck I and II telescopes and the Magellan Consortium's Landon Clay 6.5-m telescope. The observations showed that the two quasar components separated by 1\farcs41 have redshift $z=1.629$ and the lensing galaxy in between the images is at $z\sim0.3$.

\section{Photomery and image deconvolution}
\subsection{Image deconvolution}
Time delay estimation requires high precision photometry separately for each lensing image. Since our targets are small angular separation systems  we need to use a mathematical method, deconvolution, to separate the images.
\begin{figure}[!ht]
\resizebox{\hsize}{!}{\includegraphics{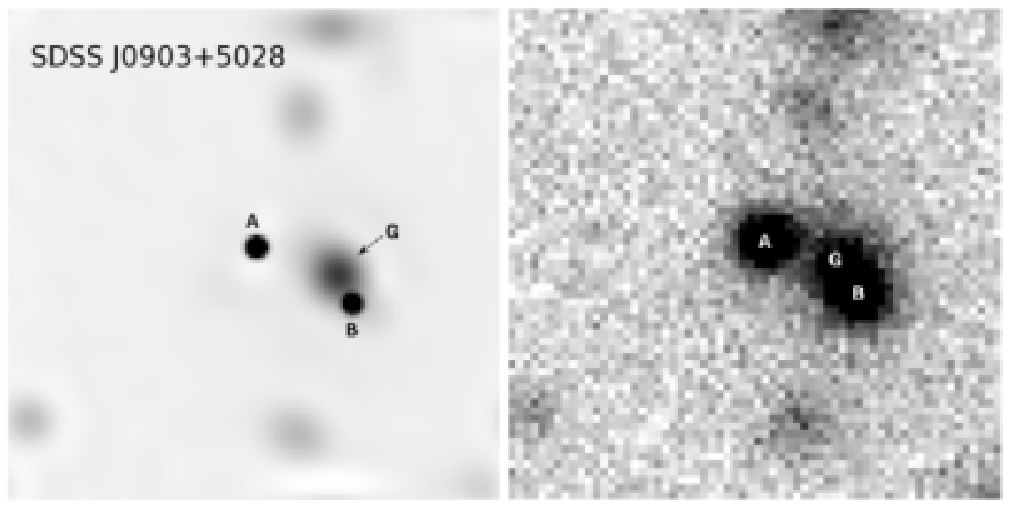}}\\
\resizebox{\hsize}{!}{\includegraphics{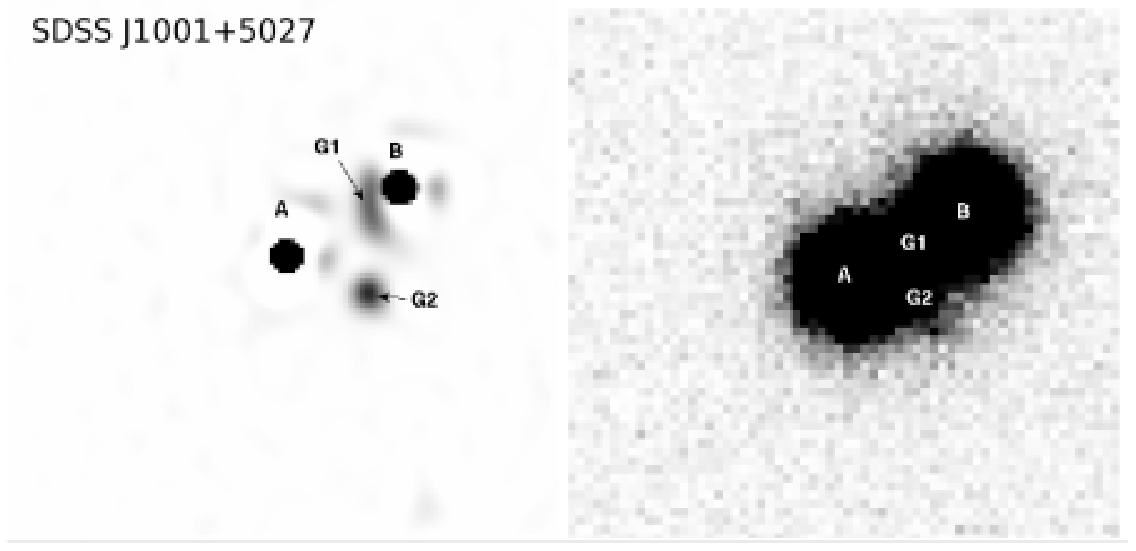}}\\
\resizebox{\hsize}{!}{\includegraphics{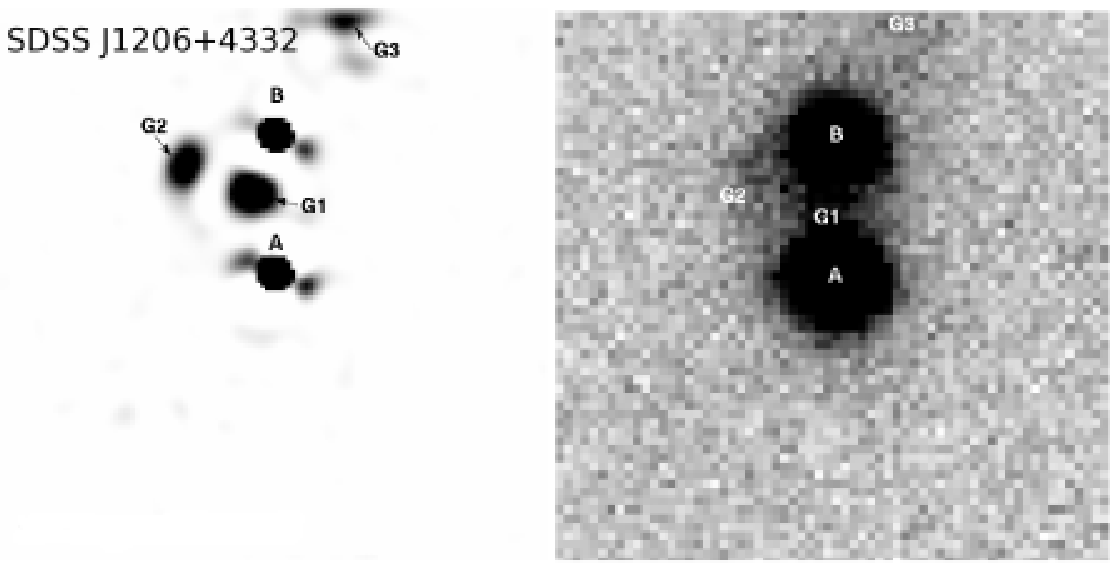}}\\
\resizebox{\hsize}{!}{\includegraphics{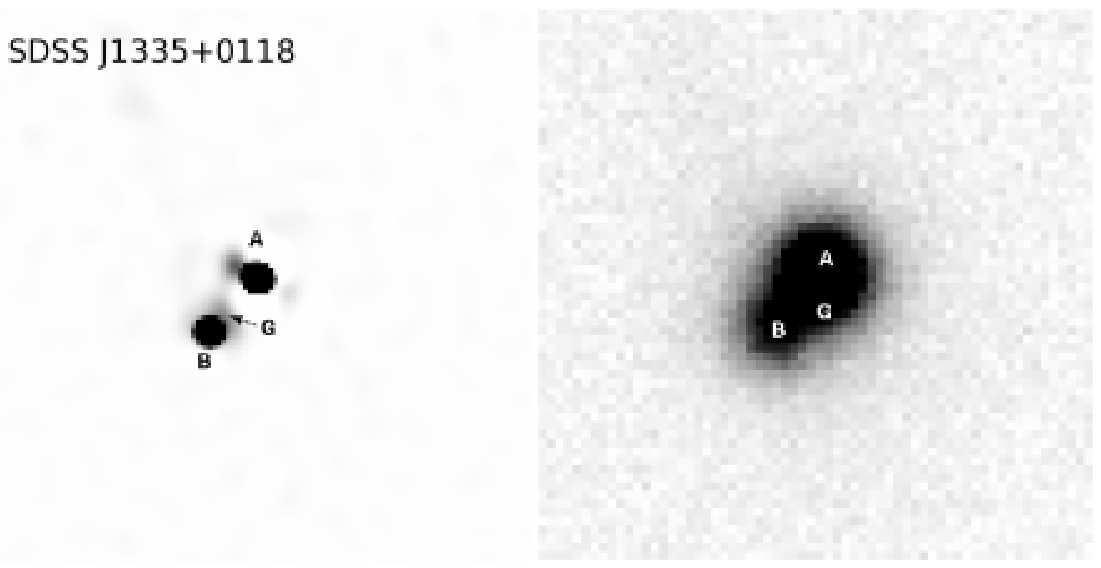}}\\
\resizebox{\hsize}{!}{\includegraphics{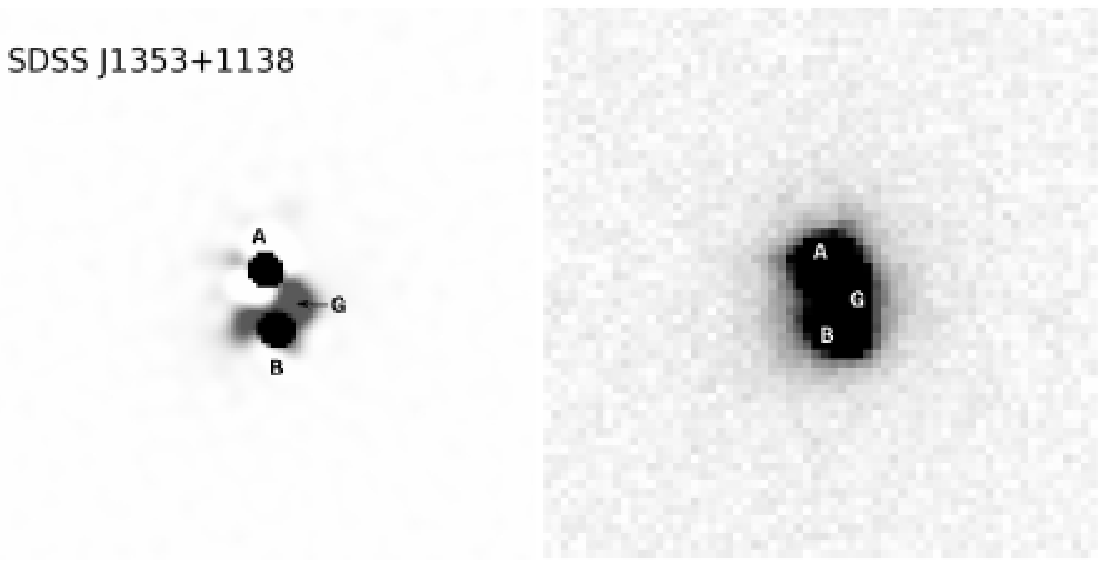}}
\caption{Results of the  deconvolution of the 5 systems. On the left are deconvolved images and on the right original images. The deconvolution reveals not only quasar images but also the lensing galaxy which is otherwise not visible.}
\label{dec}
\end{figure}

An observed image is a convolution of a real light distribution with the so-called total blurring function or point spread function (PSF).
The goal of the deconvolution is to extract the brightness of a source using knowledge about the PSF.
Deconvolution is an inverse problem without a unique solution. The method used for choosing the best solution is minimization of the difference between a model and a real image. 

All  data used in this paper were deconvolved using the PSF Controled Deconvolution software created by \citet{Magain:1998}. In this software the input are the position and intensities of the images, which need to be well def\mbox{}ined to succeed. We take the image
positions from previous studies because these were better resolved
and the quasar postions well determined (see Table \ref{phot}).

In Figure \ref{dec} we present the deconvolution results for the observed targets. On the left are shown the deconvolved images  (mathematical models of  lensing systems) and on the right the data images. We can see that the deconvolution reveals not only the quasar images but also the lensing galaxies which are not visible in the raw data.

\subsection{Photometry }
Photometry of all objects presented in this paper was made using the PSF Controled Deconvolution software \citep{Magain:1998}. The software deconvolves  all frames of a given object simultaneously, it constrains the position of the images and lensing galaxies  and allows the magnitude of the quasar images to vary freely. The algorithm has been implemented in various analysis of  quasars  \citep{Burud:2000, Burud:2002a, Burud:2002b, Hjorth:2002, Jakobsson:2004, Vuissoz:2007, Eigenbrod:2006}.  

The light curves of SDSS J0903+5028, SDSS J1335+0118,                                                                                                                                                                                                                              SDSS J1353+1138,  SDSS J1001+5027, SDSS J1206+4332 consist of 34, 31, 31, 24, 34 R-band data points, respectively, as presented in Figures 4-8. Data were taken from two optical cameras, ALFOSC (marked in black) and StanCam (marked in blue). The magnitudes are calculated relative to the PSF magnitude  marked on the f\mbox{}inding charts (Figures \ref{field1} and \ref{field2}). Due to the small f\mbox{}ield of view and comparatively small sensitivity of StanCam and lack of bright stars next to our targets  the reference stars presented in the plots are only from ALFOSC. The reference star images were also deconvolved  to look for systematic errors in the deconvolution and to estimate the photometric uncertainties. Assuming that reference stars are not intrinsically variable any variability seen in their light curves must be due to photometric and deconvolution uncertainties.  The error bars of the quasar images and the reference stars coming from ALFOSC are the averaged photon noise and uncertainties measured from the variations of the reference stars combined in quadrature. For data points  from StanCam we have assumed one standard error bar of 0.02 mag  which should approximately include all uncertainties. 
In the case of ALOSC data of SDSS J0903+5028 and StanCam data of SDSS J1335+0118 there were no non-variable star in the field except PSF star, thus the error bars of the quasar magnitude are set to 0.02 mag., which is the approximate photometric uncertainty of the B image in average atmospheric conditions.

\section{Light curves}

\subsection{QSOs with little variability}
\begin{figure}[!ht]
\resizebox{\hsize}{!}{\includegraphics{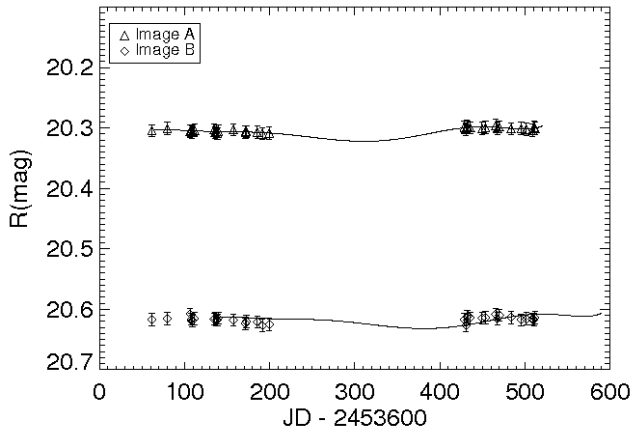}}
\caption{R-band light curves of SDSS J0903+5028. A polynomial is f\mbox{}itted to the light curve of A counterpart. This polynomial was also f\mbox{}itted to the B image with magnitude and time shift (see Table \ref{table:model_0903}). }
\label{1903}
\end{figure}

\begin{figure}[!ht]
\resizebox{\hsize}{!}{\includegraphics{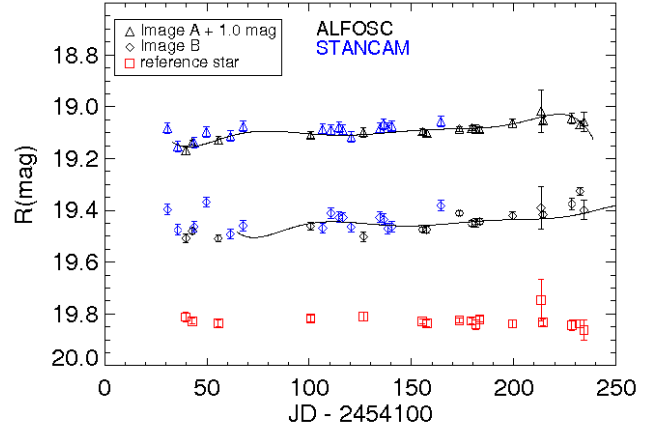}}
\caption{R-band light curves of SDSS J1335+0118. A (diamonds) and B (triangles) images of the quasar are plotted with black and blue colors indicating weather data coming from ALFOSC or StanCam. The reference star is plotted with red squares. A polynomial is f\mbox{}itted to the A counterpart of the  quasars light curve. This polynomial is also f\mbox{}itted to the B image with  magnitude and time shift. The time shift is taken from the theoretical prediction of the time delay of this system (see Table \ref{table:model_1335}).}
\label{1335}
\end{figure}

\begin{figure}[!ht]
\resizebox{\hsize}{!}{\includegraphics{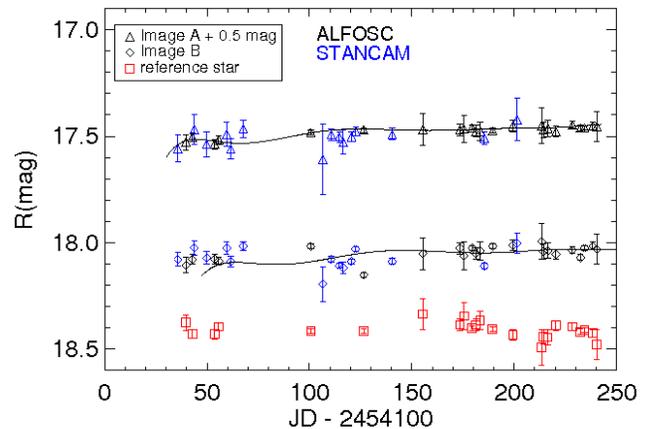}}
\caption{R-band light curves of SDSS J1353+1138. Shapes and colors of the data point and fitting procedure identical with previous light curve. The time shift is taken from the theoretical prediction of the time delay of those systems (see Table \ref{table:model_1353}). }
\label{1353}
\end{figure}

In 3 of the 5 systems (SDSS~J0903+5028, SDSS~J1335+0118, SDSS~J1353+1138) no signif\mbox{}icant variability was observed. In Figures \ref{1903}, \ref{1335} and \ref{1353}  we plot the  light curves 
of the three systems.  By visual examination we see no variability in the quasar images. 

In order to see whether small fluctuations in the light curves say something about the time delay we f\mbox{}itted polynomials to the light curves of the
A counterpart. These were then f\mbox{}itted to the B light curves with  magnitude 
and time shift. Since there are no visible peaks in either light curves, the magnitude shift was chosen simply by taking a difference between average magnitude of image A and image B.  The time shift is taken from the theoretical prediction 
of the time delay of those systems (see Section 8). 
The f\mbox{}itted polynomials do not show any preferred  time shift. This means that due to the lack of quasar variability during the time span of the monitoring we did not manage to measure  time delays in these systems.

\subsection{Quasars with variability}

In 2 out of the 5 monitored gravitationally lensed quasars variability was detected. In Figures \ref{1001} and \ref{1206} we show the light curves of SDSS J1001+5027  and SDSS J1206+4332, respectively.

\subsubsection{SDSS J1001+5027}
\begin{figure}[!ht]
\resizebox{\hsize}{!}{\includegraphics{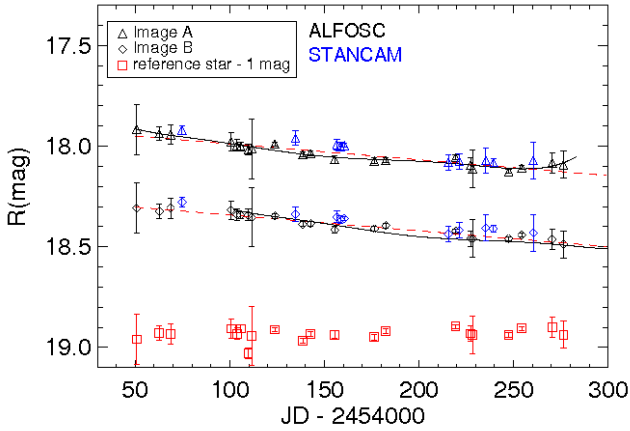}}
\caption{R-band light curves of SDSS J1001+5027. Polynomials (solid, black lines) and linear regressions (dashed, red lines) are f\mbox{}itted to the A and B counterpart of the  quasars separately. Shapes and colors of data point identical as described on previous plots.}
\label{1001}
\end{figure}

Figure \ref{1001} shows the light curves of  SDSS J1001+5027 quasar images. We see that both images have some small variabilities in the light curve and also that  both the A and B images have steadily decreased their brightness.

We have f\mbox{}itted a 7th order polynomial (see Figure \ref{1001} -  black solid line) to the A data set. This was then fitted to the B light curve with magnitude shift which is the average magnitude difference between light curves (0.4 mag) and time shift which is the predicted time delay  (52 days).

 We also have f\mbox{}itted a straight line to light curves (see Figure \ref{1001} -- red dashed lines). The linear regression shows that both light curves have very similar evolution. Both images decreased their brightness by 0.2 magnitude during the f\mbox{}irst 200 days. This indicates that the brightness decrease is due to long intrinsic quasar variability \citep{Vries:2003}.

Although variability is clearly visible,  it is impossible to measure the time delay for this system just from  the slope. The small fluctuations on the slope also do not give any conclusive results, neither visual shifting nor polynomial f\mbox{}itting  help to find the time delay.

\subsubsection{SDSS J1206+4332}
\begin{figure}[!ht]
\resizebox{\hsize}{!}{\includegraphics{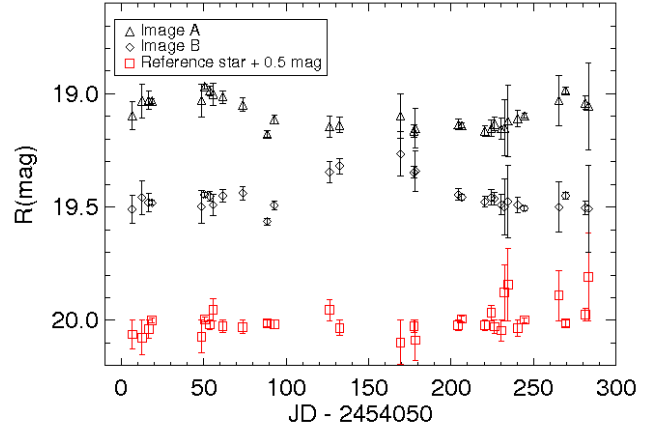}}
\caption{R-band light curves of SDSS J1206+4332.  Magnitude variabilities are seen in both light curves A (diamonds) and B (triangles) as bumps. The reference star is plotted with red squares.}
\label{1206}
\end{figure}

In Figure \ref{1206} we show the light curve of  SDSS J1206+4332. We see clear long variabilities for  both the A and B light curves. The variabilities are 100--150 days long, so the gaps in the sampling do not strongly influence the precision of the time delay estimation. The observed variabilities, consisting of bumps, allow for measurement of the time delay for SDSS J1206+4332, as detailed in
section 6.

\section{Time delay of SDSS J1206+4332}

From the geometry of the system SDSS J1206+4332 (see Figure \ref{dec}) we anticipate that the A image is  leading  since it is farther from the centre of the main lensing galaxy. Thus, we predict that any intrinsic quasar variabilities should appear f\mbox{}irst in the A image. From the mass modeling we know also that the time delay for this system is $\geq$50 days (see Section 8). We apply these constraints when measuring the time delay from the light curves.

\begin{figure}[!ht]
\resizebox{\hsize}{!}{\includegraphics{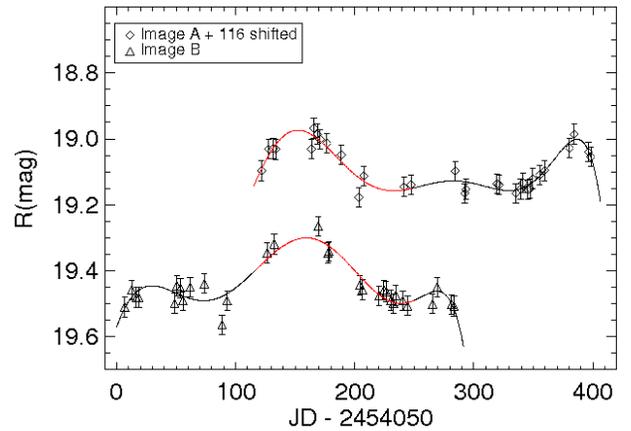}}
\caption{R-band light curves of SDSS J1206+4332.  Magnitude variabilities are seen in both light curves A (diamonds) and B (triangles) as bumps. The A light curve is shifted in time by 115 days. The red line indicates the intervals used  for cross correlation.}
\label{1206_pol}
\end{figure}

\begin{figure}[!ht]
\resizebox{\hsize}{!}{\includegraphics{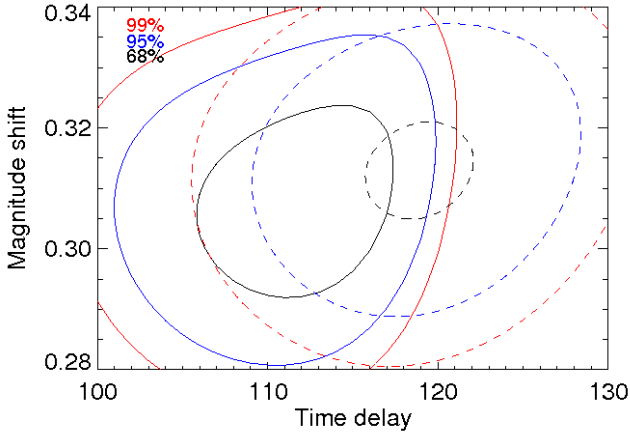}}

\caption{Results of cross correlation between data on one image  and f\mbox{}itted polynomial f\mbox{}itted to light curve of second image. Solid lines represents cross correlation between the  A data and the polynomial f\mbox{}itted to B data; dashed lines are representing cross correlation between the B data and a polynomial f\mbox{}itted to the A data. }
\label{contour}
\end{figure}

The simplest form of such a measurement is visual shifting. Knowing that the variability has to appear f\mbox{}irst in the image A and then at least 50 days later in the image B we can associate the  bump at the beginning of the A light curve centreed at $\sim$50 days with the bump in the B light curve centreed at $\sim$150 days.

A more quantitative measurement of a time delay is obtained by polynomial f\mbox{}itting. 
We f\mbox{}itted 7th order polynomials to both light curves using minimum $\chi ^2$.  
In Figure \ref{1206_pol} we show the two polynomials (solid lines).
We have marked in red the parts of the polynomials which were used for calculating the time delay. 
In order to calculate the time delay  we have  shifted the A polynomial in time and magnitude to f\mbox{}it it to the B data points and  the B polynomial was shifted in time and magnitude to f\mbox{}it it to A data points.
For each shift of the polynomials the goodness of f\mbox{}it was calculated.
The results of these f\mbox{}its are presented in Figure \ref{contour}.
For magnitude shift, $B-A = 0.31$ mag the average of the two time delay estimations is  $116^{+ 6 }_{-7}$ days, where errors are uncertainties of the two values of the time delay combined in quadrature.     Based on our modeling (G1 -- SIE and G3 -- SIS) (see Section 8) and using the measured time delay ($\Delta \tau = 116^{+ 6 }_{-7}$ days)  we find a Hubble constant of $73^{+3}_{-4}$ $\rm km\ s^{-1}\ Mpc^{-1}$.

\section{Microlensing}
\citet{Chang:1979} predicted that in lensed quasar systems the light path should be 
affected by stars in the lensing galaxy. Moving compact objects in the lensing galaxy can cause
spectral changes, brightness variability and, in the case of multiple images, flux-ratio 
anomalies in the lensed quasar. 
\begin{figure}[!ht]
\resizebox{\hsize}{!}{\includegraphics{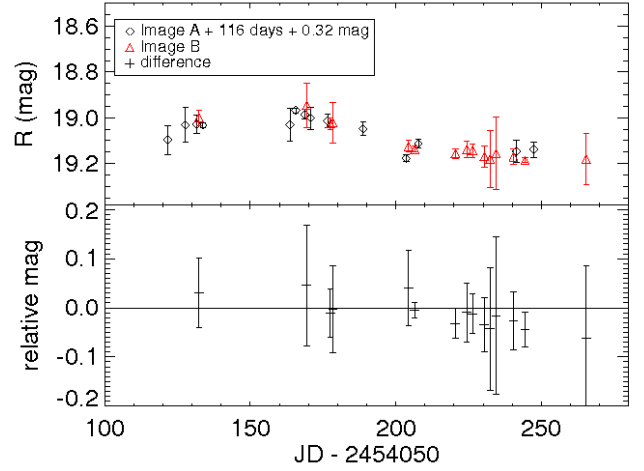}}
\caption{Microlensing  of SDSS J1206+4332. \textbf{Top:} Time-delay shifted light curves, with the A image offset by
$0.32$ mag and $115$ days.  \textbf{Bottom:} Difference between linearly interpolated A image
and B image.}
\label{microlens_1206}
\end{figure}

The images of a lensed quasar may vary due to intrinsic quasar brightness changes and/or
microlensing.
Microlensing affects the light paths of each image
differently (in the simplest case only one path is affected) whereas the
intrinsic variations show up in all the images but at different times due to
the time delay.
Therefore, one can isolate the microlensing signal by simply calculating the difference between
two light curves.

If we shift one of the light curves in magnitude (by the magnitude difference between light curves) and in time (by the time delay) and subtract it from the other quasar counterpart  light curve the remaining variation in the light curve difference should not belong intrinsically to the quasar but rather to  microlensing.

\begin{figure}[!ht]
\resizebox{\hsize}{!}{\includegraphics{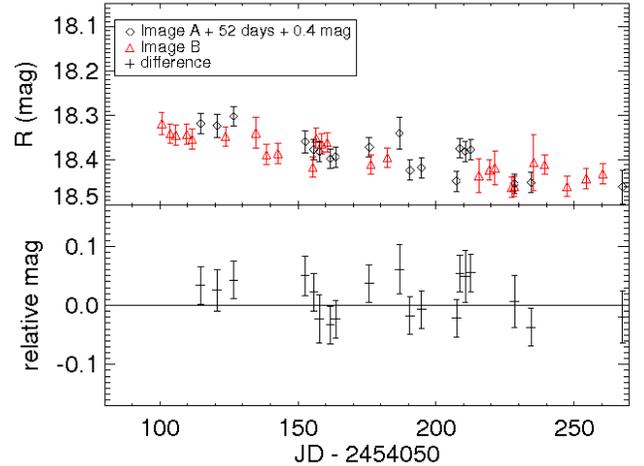}}
\caption{Microlensing of SDSS J1001+5027. \textbf{Top:} Time-delay shifted light curves, with the A image offset by
$0.4$ mag and $52$ days.  \textbf{Bottom:} Difference between linearly interpolated A image and B image.}

\label{microlens_1001}
\end{figure}

\begin{figure}[!ht]
\resizebox{\hsize}{!}{\includegraphics{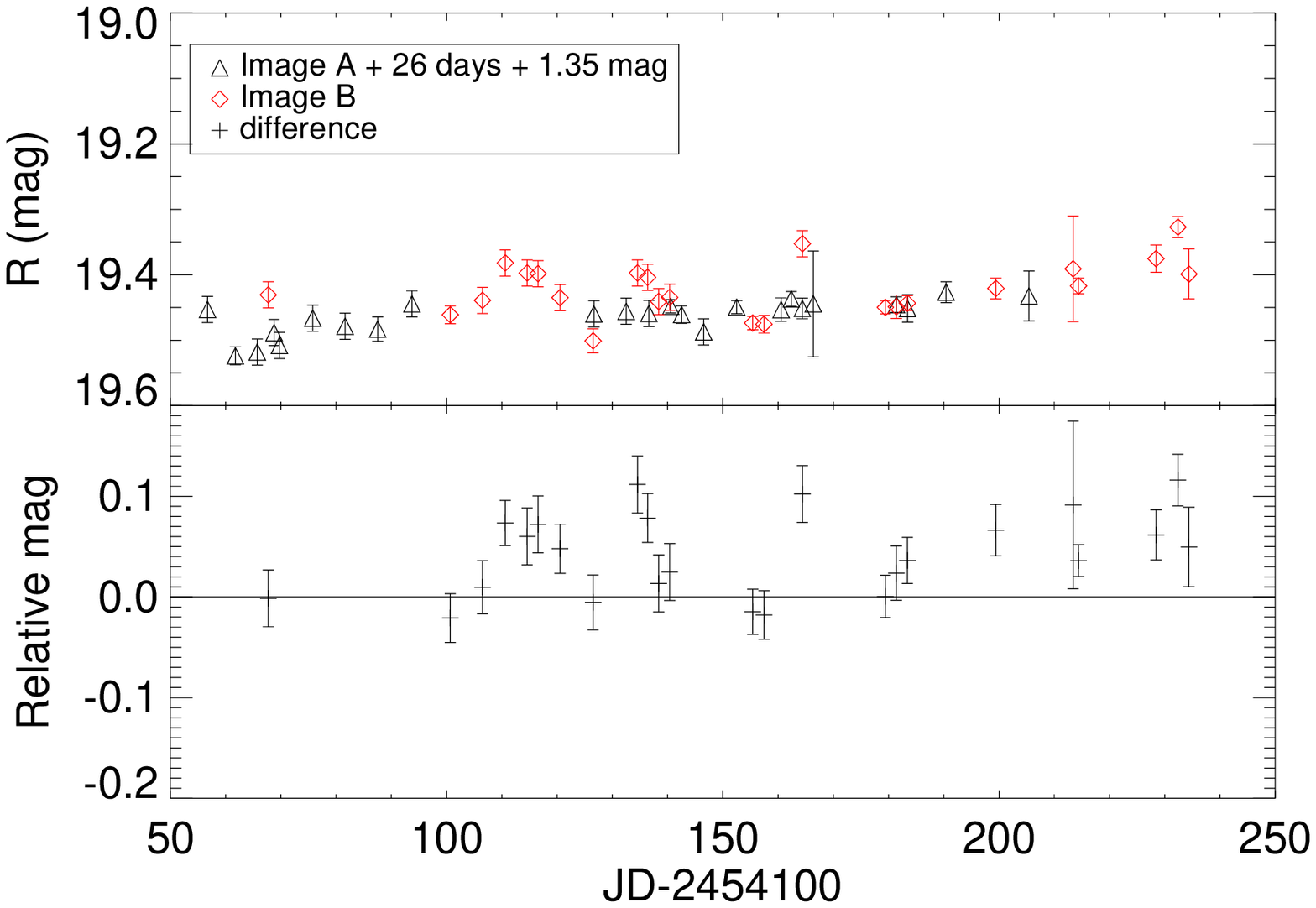}}
\caption{Microlensing of SDSS J1335+0118. \textbf{Top:} Time-delay shifted light curves, with the A image offset by
$1.35$ mag and $26$ days.  \textbf{Bottom:} Difference between linearly interpolated A image and B image.}
\label{microlens_1335}
\end{figure}

\begin{figure}[!ht]
\resizebox{\hsize}{!}{\includegraphics{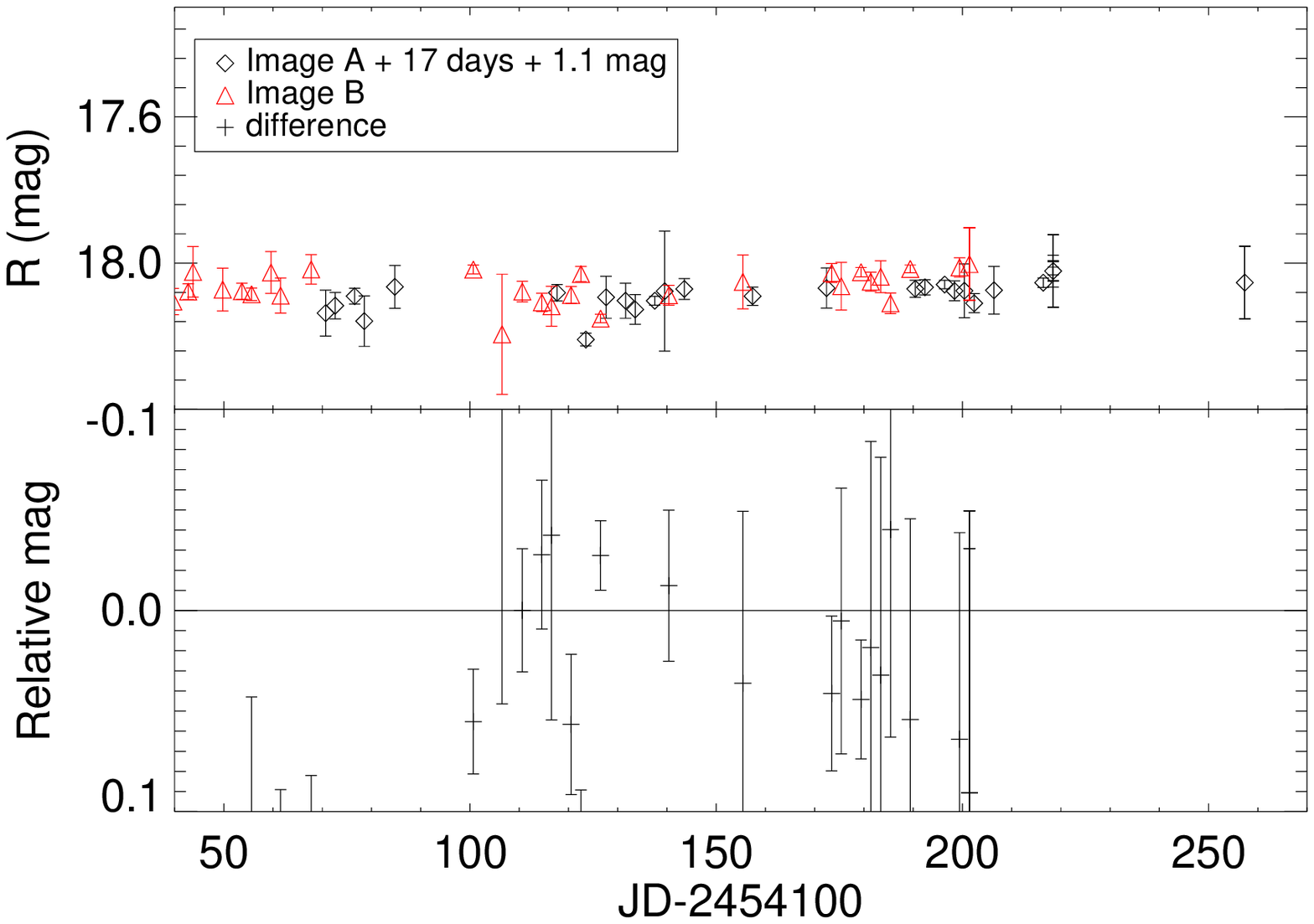}}
\caption{Microlensing of SDSS J1353+1138. \textbf{Top:} Time-delay shifted light curves, with the A image offset by
$1.1$ mag and $17$ days.  \textbf{Bottom:} Difference between linearly interpolated A image and B image.}
\label{microlens_1353}
\end{figure}

\begin{figure}[!ht]
\resizebox{\hsize}{!}{\includegraphics{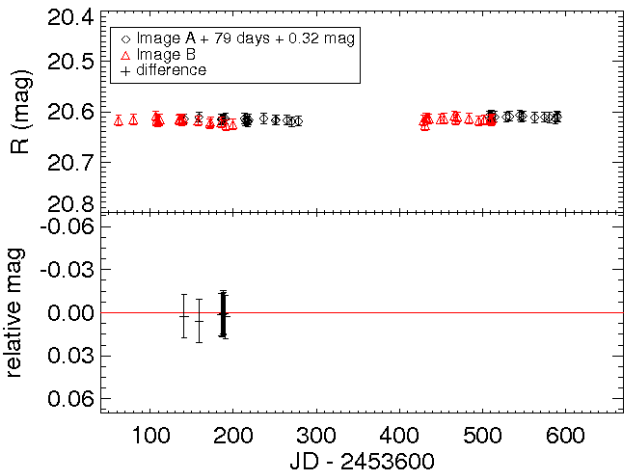}}
\caption{Microlensing of SDSS J0903+5028. \textbf{Top:} Time-delay shifted light curves, with the A image offset by
$0.32$ mag and $79$ days. \textbf{Bottom:} Difference between linearly interpolated A image and B image.}
\label{microlens_1903}
\end{figure}

Our previous study \citep{Paraficz:2006} showed that microlensing variabiliy
can be detected in this way. The monitoring program presented in this paper was designed to be sensitive to such microlensing signals. The only limitation are S/N and gaps in the light curve due to bad weather conditions. In Figures 11--15 we present the result of a microlensing variability analysis for our observed targets. 
                      
The top plots present the quasar image lightcurves, with one of the images being shifted in 
time (by the  theoretically predicted or, if available, the measured time delay) and in magnitude 
(by the average magnitude difference between the images). The lower panels show the 
light curve differences. The difference is calculated  between a  linearly interpolated  B light curve and the A data points.
In Figures  11--15 we see that there is no signal stronger than 0.1 mag which is $\sim$ 2 $\sigma$. 
We therefore conclude that  we did not detect any  significant microlensing event
in any of the systems during our monitoring program.

\section{Modeling} 
\subsection{Analytical modeling}

In principle, one can calculate the Hubble constant from
the time delay of a gravitationally lensed system,
but there is one major obstacle, namely the mass-sheet degeneracy 
\citep{Falco:1985}. This degeneracy is between the Hubble constant and 
the radial density profiles of the lens \citep{Wucknitz:2001} and its angular 
structure \citep{Freedman:2001}. Summarizing, different mass profiles give 
different results on the Hubble constant without changing the lensing
configuration. To break this degeneracy we need to have information about 
the mass profile, e.g., through the velocity dispersion 
 \citep{Romanowsky:1999}. For our  systems we do not know which lens 
model is the correct one, but fortunately in many cases a lens can be quite 
well modeled using  a fairly simple mass model like SIE 
(Singular  Isothermal Ellipsoid)  \citep{Kormann:1994}  or NFW 
(Navarro-Frank-White) \citep{Golse:2002}.

In this paper, for the analytical modeling,  we have modeled all our systems using singular isothermal potentials. We used those models because of their simplicity and because they agree with the 
physical properties of many observed lenses  \citep{Kochanek:1993, Rusin:2003, Koopmans:2006}. 

In the analytical modeling we have the positions of the two images as constraints and as free parameters  we have the lens velocity dispersion, $\sigma_0$ and the ellipticity, $\epsilon$. We also have two fixed parameters,  the position angle, $\theta_{\epsilon}$ and the position of the lens galaxy. The number of degrees of freedom is 0, thus we can look for the one model which perfectly fits the data.
The positions of the images and lens are visible on the 
decovolved images (see Figure \ref{dec}). 
The position angle of the lenses can be roughly estimated from analysis of the deconvoled images,  as the mass profile of a lens tends to align to its visible component  \citep{Keeton:1998}. We allow this constraint to vary within $45^{\circ}$.  We have not used  flux ratios as constraints because of the possible influence of reddening by dust \citep{Eliasdottir:2006}, microlensing \citep{Paraficz:2006} or small-scale structure in the lens potential \citep{Dalal:2002}.

The ellipticity used here is defined as 
$\epsilon = (a^{2}-b^{2})/(a^{2}+b^{2})$, where a and b are the major and minor axis. The  position angle  corresponds
to the direction of the semi-major axis of the isopotential counted east of north.
The modeling has been performed
using the LENSTOOL software package 
available at  http://www.oamp.fr/cosmology/lenstool/ \citep{Jullo:2007}. 
LENSTOOL is a software created for modelling strong lensing systems with 
parametric methods which chooses models using the Bayesian evidence. 
In all models we set the Hubble constant to be $H_0=73$ $\rm km\ s^{-1}\ Mpc^{-1}$.
\subsubsection{SDSS J0903+5028}
\begin{figure}
\centering
\resizebox{\hsize}{!}{\includegraphics{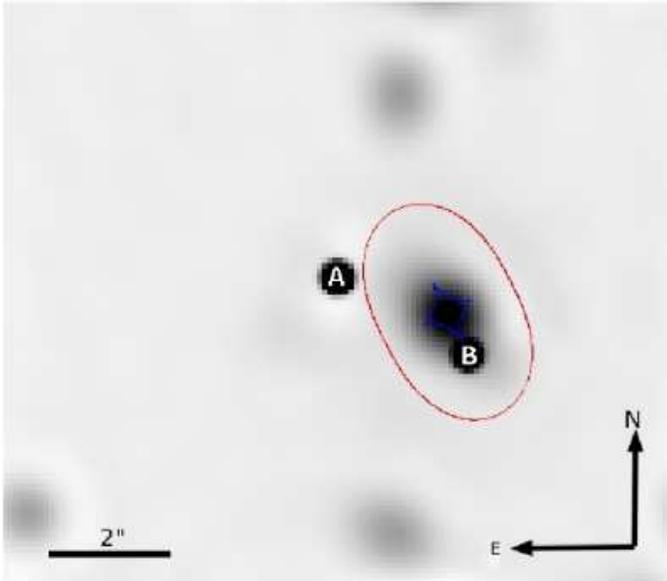}}
   \caption{Lens model of SDSS J0903+5028. The figure shows an image of the system with over-plotted results of lens modeling. Blue and red lines represent critical and caustic lines, respectively.}
    \label{contour0903}
\end{figure}

We have fitted the two simple models, SIS (Singular Isothermal Sphere) and SIE 
(Singular Isothermal Ellipsoid)
to the lensing galaxy of SDSS J0903+5028. 
We have assumed that only one galaxy, the central one, influences the quasar light,  even though other galaxies are visible to the north and south of the images (see Figure \ref{contour0903}). The lensing images of the quasar are not collinear with the central galaxy which indicates  that there is a quadruple moment in the potential. This moment can come from the tidal effect of the nearby galaxies or may be due to elongation of the central galaxy. Because of the degeneracy we are not able to distinguish which one of the possibilities is true,  thus we use one lens, an SIE model which accurately predicts image positions. In Table \ref{table:model_0903} we summarize the results of the SDSS J0903+5028 modeling where the central single lens is an SIE.   Based on this model, the predicted time delay for the system is 79.4 days. The uncertainty in this value is
entirely dominated by systematic modeling uncertainties.

\begin{table}[!ht]

\centering
\begin{tabular}{c c c c c c c c}
\hline \hline
\#&Type&x&y  &$\sigma_0$ &$\epsilon$ & $\theta_{\epsilon}$&$\Delta\tau$ \\
& &arcsec&arcsec&$\rm km\ s^{-1}$ & & &days \\
\hline
G&SIE&1.99&-0.70& 264.6& 0.49 &$27^{\circ}$&79.4\\ 
\hline
\end{tabular}
\caption{Lens model of SDSS J0903+5028. Modeled parameters are galaxy position, $x,y$, lens velocity dispersion, $\sigma_0$, ellipticity, $\epsilon$, position angle, $\theta_{\epsilon}$ and time delay, $\Delta\tau$.}
\label{table:model_0903}
\end{table}

\subsubsection{SDSS J1001+5027} 
The lens of SDSS J1001+5027 consists of two galaxies 
(see Figure \ref{dec}) which according \citet{Oguri:2005} are at similar 
redshifts $0.2<z<0.5$. For our modeling we fix both lensing galaxies at redshift 0.3.

\begin{figure}
\centering
\resizebox{\hsize}{!}{\includegraphics{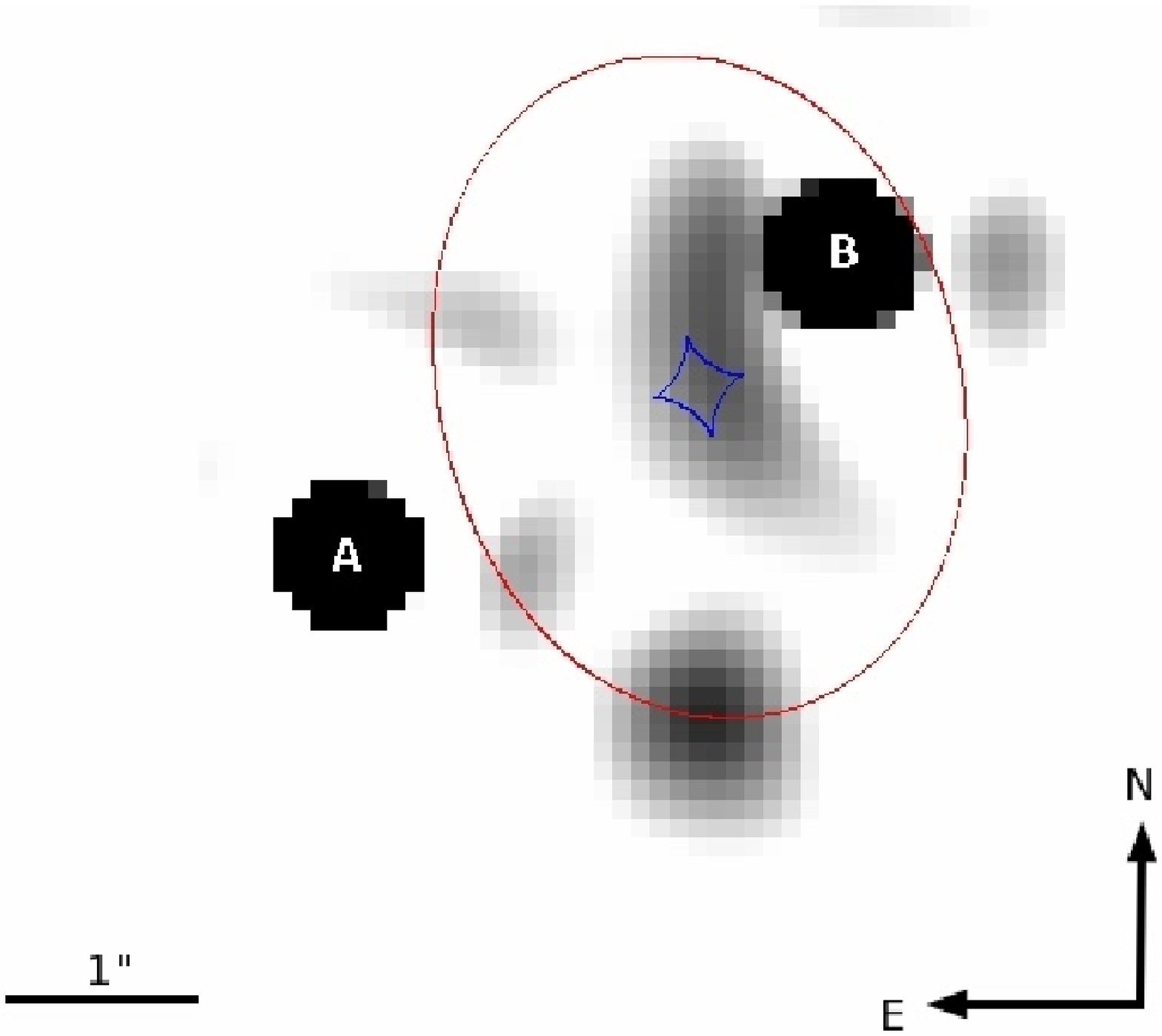}}
\resizebox{\hsize}{!}{\includegraphics{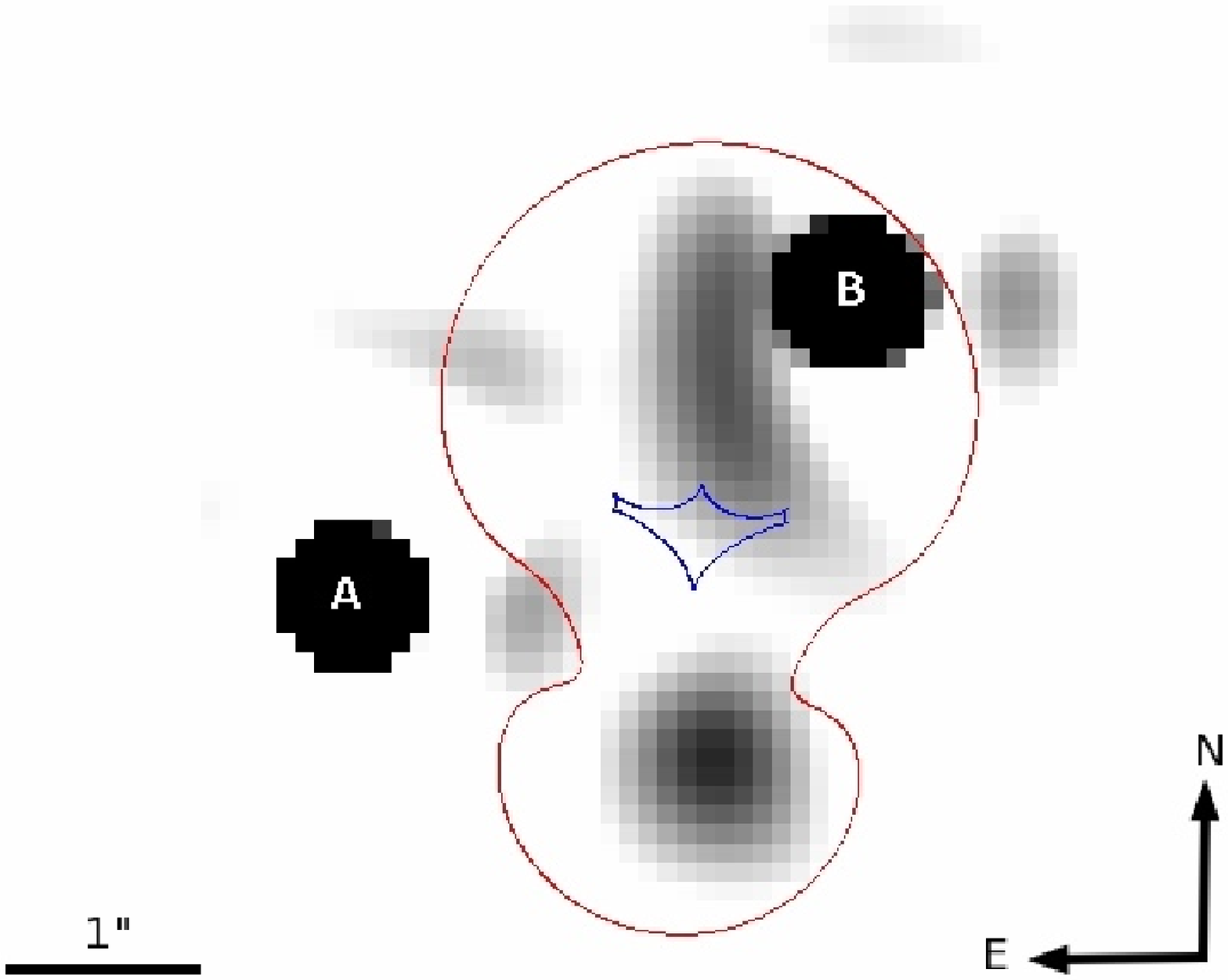}}
    \caption{Lens model of SDSS J1001+5027. The figure shows an image of the system with over-plotted results of lens modeling. Blue and red lines represent critical and caustic lines, respectively. Top plot present the single SIE lens model and the bottom plot presents the two SIS lens models.}
    \label{contour1001}
\end{figure}

We have created two possible models of the lens environment of SDSS~J1001+5027. 
In the first one the system has only one galaxy with elongated SIE mass profile.
This model  is able to ray-trace the position of the quasar images. From 
Figure \ref{contour1001} we also see that  the position angle, P.A. of the model is consistent with the light distribution of the main galaxy but it is also in the direction of the other galaxy. This leads to the conclusion 
that the quadruple moment might come from the second galaxy rather then from
the elongation of the main one. In the second version of the lens model we have set both gravitational lenses as Singular Isothermal Spheres, SIS. 
This model also reproduces the position of the quasar images and gives a sensible result of the galaxy velocity dispersions being 223 $\rm km\ s^{-1}$ for main lens (G1) and 159 $\rm km\ s^{-1}$ for G2.
The results from both models are presented in Figure \ref{contour1001} 
and summarized in Table \ref{table:model_1001}. 
The predicted time delay of the system is 34 days based on the SIE model and 52 days based on the double SIS model.

\begin{table}[!ht]
\centering
\begin{tabular}{c c c c c c c c}
\hline\hline
\#&Type&x&y    &$\sigma_0$ &$\epsilon$ & $\theta_{\epsilon}$&$\Delta\tau$ \\
& &arcsec&arcsec&$\rm km\ s^{-1}$  & & &days \\
\hline
G1&SIE&1.749&0.861& 258& 0.23 &$11^{\circ}$&52\\ 
\hline
G1&SIS&1.749&0.861& 223&0&0&34\\
G2&SIS&1.629&$-0.588$& 159&0&0&34\\
\hline
\end{tabular}
\caption{Lens model of SDSS J1001+5027. Parameters are as  in Table~\ref{table:model_0903}.}
\label{table:model_1001}
\end{table}

\subsubsection{SDSS J1206+4332}
We have fitted SIS and SIE to the main lensing galaxy of the SDSS J1206+4332.
We have assumed that the only galaxy which influences the quasar light is the 
central galaxy G1 (see Figure \ref{dec}).
Unfortunately  we come to the same conclusions as \citep{Oguri:2005} that neither of these mass models gives physically realistic results.  
The system seems to be more complicated, so presumably the other galaxies seen 
in vicinity of the system may play a role. A model of the main galaxy G1 
(SIE) and G2 (SIS) does not give any physical solutions either, which 
is expected since the redshift of G2 is predicted to be rather small. 
 
It appears that the third galaxy lying $\sim3\arcsec$ north of the system 
(see Figure \ref{dec}) has a strong  influence on the system geometry. 
If we assume our model to be two lensing galaxies with one of them being G1 
(SIE) and the other G3 (SIS) it is possible to reproduce the positions
of the images. The results of that final fitting are summarized in Table 
\ref{table:model_1206}. 
Based on this model, the predicted time delay of the system is 112.5 days 
which is in agreement with our measurement of $116^{+ 6 }_{-7}$ days (see section 6).

\begin{figure}
\centering
\resizebox{\hsize}{!}{\includegraphics{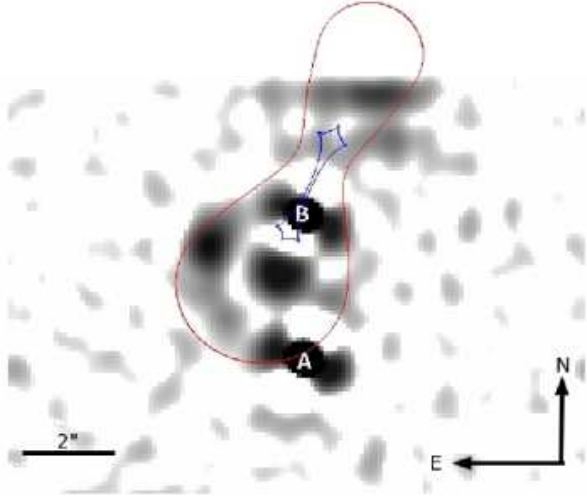}}
   \caption{Lens model of SDSS J1206+4332. The figure shows an image of the 
   system with over-plotted results of lens modeling. Blue and red lines represent critical and caustic lines respectively.}
    \label{contour1206}
\end{figure}

\begin{table}[!ht]
\centering
\begin{tabular}{c c c c c c c c}

\hline\hline 
\#&Type &x&y &$\sigma_0$ &$\epsilon$ & $\theta_{\epsilon}$&$\Delta\tau$ \\
& &arcsec&arcsec&$\rm km\ s^{-1}$  & & &days \\
\hline    
G1&SIE&$-0.66$&1.75& 347.5& 0.1 &$-2^{\circ}$&112.5\\ 
G3&SIS& 1.32&5.9 &279.5& 0& -&112.5\\

\hline
\end{tabular}
\caption{Lens model of SDSS J1206+4332. Parameters are as  in Table~\ref{table:model_0903}}
\label{table:model_1206}
\end{table}

\subsubsection{SDSS J1335+0118}

The system has one lensing galaxy, at 
redshift $z=0.5$, which we have modeled as an SIE. The lensing galaxy is barely visible in our images, hence
the position of the galaxy was taken from \citet{Oguri:2004}. According to
observations made by \citet{Oguri:2004} with Keck, the lensing galaxy is 
misaligned with the QSO images which indicates the existence of external 
shear or elongation of the lensing galaxy.
A model with an SIE lens accurately predicts the image position and gives
plausible galaxy parameters.
 The model details are summarized in Table \ref{table:model_1335}. The 
 estimated time delay of the system is 26.2 days
 which is in agreement with previous studies. 

\begin{figure}
\resizebox{\hsize}{!}{\includegraphics{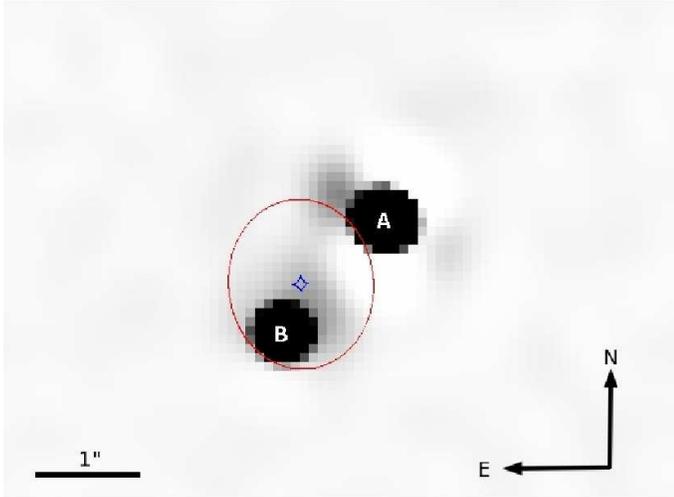}}
\caption{Mass model of SDSS J1335+0118. Figure shows an image of the system with over-plotted results of lens modeling. Blue and red lines represent critical and caustic lines, respectively.}
\label{contour_1335}
\end{figure}
\begin{figure}
\resizebox{\hsize}{!}{\includegraphics{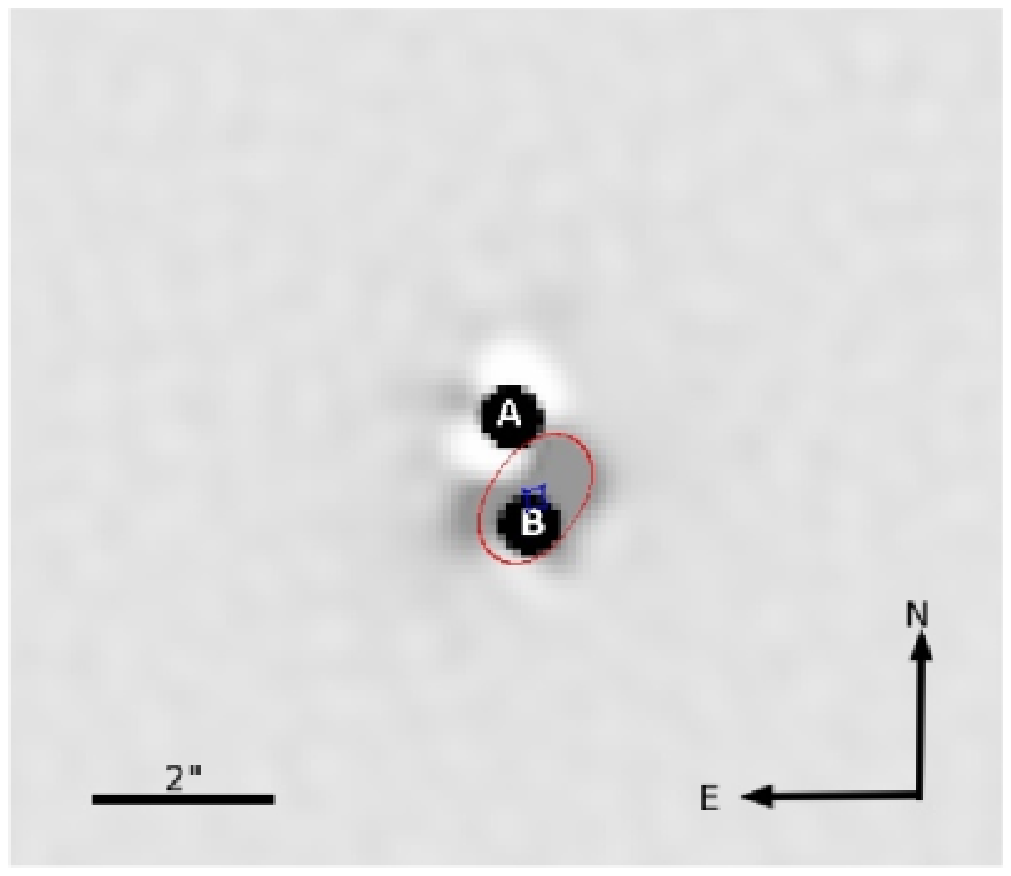}}
\caption{Mass model of SDSS J1353+1138. Figure shows an image of the system with over-plotted results of lens modeling. Blue and red lines represent critical and caustic lines, respectively.}
\label{model}
\end{figure}

\begin{table}[!ht]

\centering
\begin{tabular}{c c c c c c c c}
\hline\hline
\#&Type&x&y&$\sigma_0$ &$\epsilon$ & $\theta_{\epsilon}$ &$\Delta\tau$\\
& &arcsec&arcsec&$\rm km\ s^{-1}$  & & &days \\
\hline
 G&SIE&$-0.786$&$-0.649$& 218& 0.16 &$-5^{\circ}$&26.2\\ 
\hline
\end{tabular}
\caption{Lens model of SDSS J1335+0118. Parameters are as  in Table~\ref{table:model_0903}}
\label{table:model_1335}
\end{table}

\subsubsection{SDSS J1353+1138}
 
  SDSS J1353+1138 has a centrally aligned lensing galaxy and the B image very close to the galaxy centre.  There are no other close-by galaxies; thus the lens can be modeled with a simple SIE. The predicted time delay for that system is 17 days (see Table \ref{table:model_1353}) which is in agreement with previous studies.

\begin{table}[!ht]

\centering
\begin{tabular}{c c c c c c c c}
\hline\hline
\#&Type&x&y  &$\sigma_0$ &$\epsilon$ & $\theta_{\epsilon}$&$\Delta\tau$ \\
& &arcsec&arcsec&$\rm km\ s^{-1}$  & & &days \\
\hline
G&SIE&$-0.25$&$-1.04$& 177& 0.46&$35^{\circ}$&17.2\\ 
\hline
\end{tabular}
\caption{Lens model of SDSS J1353+1138. Parameters are as  in Table~\ref{table:model_0903}}
\label{table:model_1353}
\end{table}

\subsection{Non-parametric modeling}

We have also modeled our systems based on a non-parametric method proposed by \citet{Saha:1997} using the publicly available code PixeLens \citep{Saha:2004}.

PixeLens reconstructs a pixelated mass map of the lens by implementing Bayesian statistics. It generates numerous lens models which fit the lens system geometry.
There are two major advantages of using this method. First, it allows to explore a wide range of models not restricted with parameters, which circumvents the non-uniqueness problem of lens modeling. Second, it provides an estimate of the systematic errors on the modeled time delay which is not possible to get when using analytical methods.  
Our purpose of using it is to check the robustness of our time delay estimations, i.e. we want to  know whether the two approaches, analytical and pixeleted, give consistent results for the time delays.
\begin{figure}[!htpb]
\resizebox{\hsize}{!}{\includegraphics{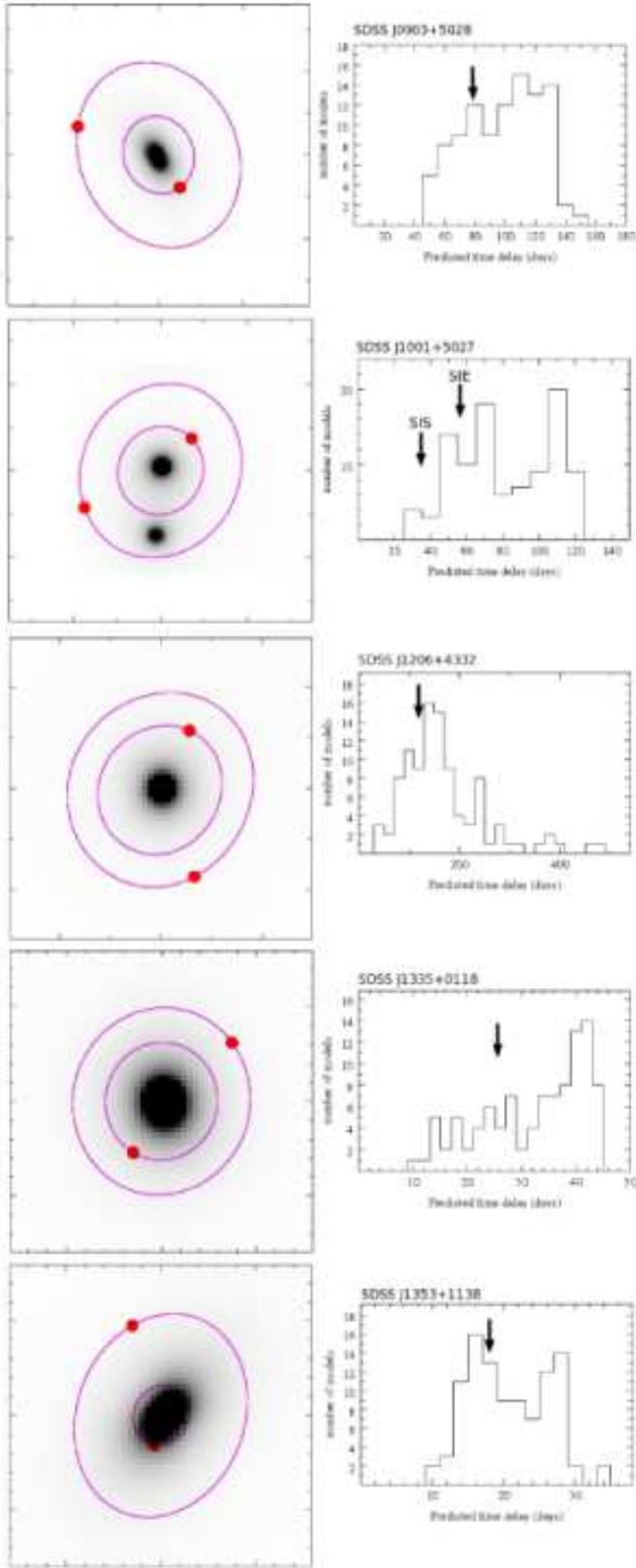}}
\caption{The modeling  results of the five lensed quasars. On the left side we see potential contours modeled by PixeLens which are  over-plotted on the mass map created in analytical modeling. On the right the are time delay histograms generated for each system by the 300 suitable lens models, arrows indicate the time delays estimated from analytical modeling.
We see that the potentials of the PixeLens correspond quite well to the analytical models}
\label{model_all_poten}
\end{figure}
PixeLens  generates models which reproduce the exact position of lensing images and aligned within  45$^\circ$ of the visible lens P.A. All our systems were modeled assuming existence of external shear and all of them, except SDSS J1206+4332, were modeled as symmetric. For each system we generated 300 lens models which were used to estimate the average and median time delay  and their model uncertainties. The modeling results of the 5 gravitationally lensed QSOs with PixeLens are summarized in Table \ref{table:alldelay}.  
Figure \ref{model_all_poten} shows the potentials and time delays of the 5 systems modeled with pixelated method. On the left side we see potential contours modeled by PixeLens which are over-plotted on the mass map created in analytical modeling. On the right are histograms  of the time delays generated for each system by the 300 lens models.

We see that the lens potentials of the PixeLens correspond to the analytical models and from the Table \ref{table:alldelay} we see that the estimated time delay values from analytical modeling lie within error bars of  the PixeLens time delays.  This is reassuring and gives confidence in our approach. We note also that the analytically obtained time delays are systematically lower than the average values obtained using PixeLens by 20--50\%.

\begin{table}[!ht]
\centering
\begin{tabular}{c c c c }
\hline\hline
\#&Analytical&PixeLens&PixeLens \\
&Model &average&median \\
\hline
0903+5028&79.4&98$\pm$21&$101^{+9}_{-12}$\\ 
1001+5027&52&81$\pm$23&$76^{+14}_{-9}$\\
1206+4332&113&173$\pm64$&$152^{+41}_{-21}$\\
1335+0118&26&32$\pm$8&$35^{+2}_{-6}$\\ 
1353+1138&17&21$\pm5$&$20^{+3}_{-2}$\\ 
\hline
\end{tabular}
\caption{Summary of modeled time delays of the five lensed quasars. Two different modeling methods are compared. For the pixelated method  we generate 300 lens models for each system which are used to estimate the average and median time delay  and their error bars. We see that the estimated time delays from analytical modeling are consistent with he PixeLens time delays.}
\label{table:alldelay}
\end{table}

\section{Simultanous modeling}

The Hubble constant estimated from time delay measurements is model dependent, as illustrated above.  Different models give a different Hubble constant and measuring the time delay with high precision does not resolve that issue. 
One way of dealing with this problem is to generate a large number of models of many lenses and create a distribution of possible values for the Hubble constant based on these models. Thus, we use the pixelated method for several gravitationally lensed systems constraining them to have the same shared value of $H_0$.
Simultaneous modeling has been done in the past by e.g. \citet{Saha:2006} -- with 10 lenses and \citet{Coles:2008} -- with 11 lenses. 

\begin{figure}[!htpb]
\resizebox{\hsize}{!}{\includegraphics{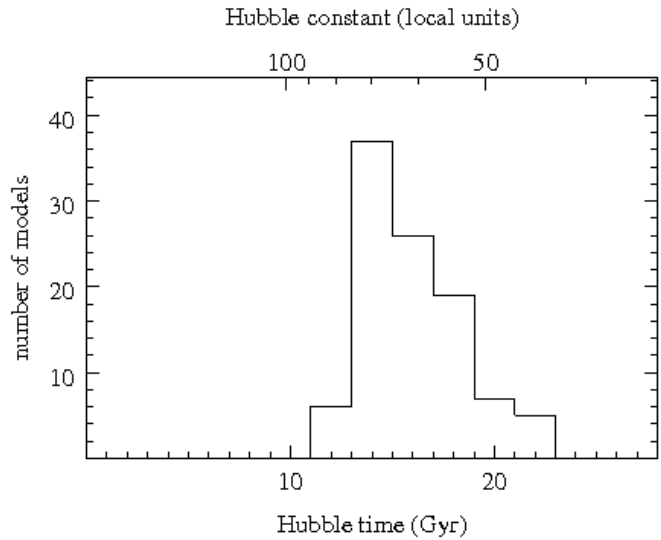}}
\caption{The result of the simultaneous modeling, it is a histogram created from the modeling of the 5 QSOs monitored at the NOT for each of them 100 lens models were created, the Hubble constant estimated using this method is $ H_0 =61.5^{+8}_{-4}$ $\rm km\ s^{-1}\ Mpc^{-1}$.}
\label{model_all_delay}
\end{figure}

Figure \ref{model_all_delay} shows the result of the simultaneous modeling of
5 systems with time delays obtained at the NOT. These are RX J0911+055 \citep{Hjorth:2002}, SBS B1520+530, \citep{Burud:2002b}, B1600+434 \citep{Burud:2000} and J0951+263 \citep{Jakobsson:2004} and the time delay of SDSS J1206+4332 reported in this paper. On the histogram the results of our simultaneous modeling of the 5 QSOs are presented. For each object 100 lens models were created. The Hubble constant estimated using this method is $H_0 =61.5^{+8}_{-4}$ $\rm km\ s^{-1}\ Mpc^{-1}$.

\section{Discussion}

We have estimated the time delay of the  SDSS J1206+4332 to be $\Delta \tau = 116^{+ 6 }_{-7}$ based on a 280-day long monitoring campaign at the Nordic Optical Telescope. This shows the feasibility of measuring time delays from campaigns not much longer than the time delay itself.
To model the system, we first assumed a single lensing galaxy modeled with the Singular Isothermal Ellipsoid (SIE), but were unable to reproduce the image positions.
By adding to the lens model the G2 and G3 galaxies we found that the influence of the G2 galaxy is negligible due to its small redshift  but the G3 galaxy has a major impact on the system's geometry.  A lens model with two galaxies, where the main one is a SIE and the second is a SIS, predicts the image positions very well. Thus this model 
  resulted in a Hubble constant of $73^{+3}_{-4}$ $\rm km\ s^{-1}\ Mpc^{-1}$ assuming  $\Omega_{\rm m}=0.3$ and $\Omega_{\Lambda}=0.7$. 

Lack of short-term variability in the other monitored systems (SDSSJ 0903, SDSS J1001+5027,  SDSS J1353+1138, SDSS J1335+0118) meant that  we did not succeed in measuring their time delays.  From the quasar studies of \citet{Fohlmeister:2008}, \citet{Vuissoz:2007}, \citet{Koopmans:2000} and many others we see that during the 200--300 days of a quasar monitoring a quasar might be at the quiet stage or at the slow increase or decrease. Thus, despite our success in measuring the time delay of SDSS J1206+4332, it appears that monitoring substantially longer than the predicted time delay of a given system will be unavoidable in many
cases.

We have modeled all five systems using both analytical and pixelated methods. Using the predicted time delays we have shown that none of the systems exhibited
significant microlensing variability during the observing campaign.

We have also performed simultaneous pixelated modeling with a common Hubble constant of five lenses for which time delays were measured at the NOT. The 
estimated Hubble constant from this analysis is 
$H_0 =61.5^{+8}_{-4}$ $\rm km\ s^{-1}\ Mpc^{-1}$.

\textit{Acknowledgments}
The Dark Cosmology Centre is funded by the DNRF.
DP acknowledges receipt of a research studentship at the Nordic Optical Telescope.  \'A.E. acknowledges the support of the EU under a Marie Curie International Outgoing Fellowship, contract PIOF-GA-2008-220049.
We thank Andreas Jaunsen for his contribution in the definition phase of the project,
Jin Hyeok An and Andrea Morandi for useful comments and P\'all Jakobsson for extensive help with  MCS deconvolution. Also, we would like to thank all visiting observers at the NOT who obtained data on our behalf.
The data used are based on observations made with the Nordic Optical Telescope, operated
on the island of La Palma jointly by Denmark, Finland, Iceland,
Norway, and Sweden, in the Spanish Observatorio del Roque de los
Muchachos of the Instituto de Astrof\mbox{}isica de Canarias.

\newpage
\appendix
\scriptsize

\begin{center}
\begin{table*}
\centering

{\bf Table.3.} Photometry of  two images of SDSS J0903+5028 quasar.

\centering
\begin{tabular}{|c|c|c|}
\hline
$Rmag_A$&$Rmag_B$&JD\\
\hline
20.304195 & 20.616799 & 061.73267\\
20.300746 & 20.615271 & 079.76126\\
20.305079 & 20.608090 & 106.76752\\
20.304733 & 20.616682 & 107.72829\\
20.306558 & 20.619706 & 108.77146\\
20.304743 & 20.618817 & 109.75729\\
20.303631 & 20.615568 & 111.70990\\
20.302910 & 20.616263 & 134.49363\\
20.307745 & 20.615292 & 135.60198\\
20.305574 & 20.617903 & 137.56633\\
20.309453 & 20.618127 & 138.65267\\
20.305457 & 20.614339 & 139.65707\\
20.302961 & 20.618014 & 157.56273\\
20.306617 & 20.623384 & 171.54778\\
20.305926 & 20.620051 & 172.58139\\
20.306603 & 20.620899 & 185.40632\\
20.309605 & 20.626972 & 191.46353\\
20.308627 & 20.625028 & 199.52655\\
20.298319 & 20.617214 & 428.69561\\
20.301808 & 20.627140 & 430.74676\\
20.296764 & 20.612383 & 431.69420\\
20.298253 & 20.612072 & 432.73686\\
20.299821 & 20.614314 & 435.74226\\
20.300707 & 20.614775 & 449.59337\\
20.299069 & 20.613606 & 453.71216\\
20.296104 & 20.608069 & 465.70068\\
20.300696 & 20.615813 & 467.67382\\
20.298169 & 20.609918 & 469.74747\\
20.301014 & 20.613560 & 483.76553\\
20.300752 & 20.617619 & 495.56600\\
20.301731 & 20.615859 & 501.74698\\
20.303775 & 20.616146 & 508.74511\\
\hline
\end{tabular}
\end{table*}
\end{center}

\scriptsize
\begin{center}

\begin{table*}
\centering
{\bf Table.3.} Photometry of  two images of SDSS J1001+5027 quasar and a reference star.
\centering  

\begin{tabular}{|c|c|c|c|c|}
\hline
$Rmag_A$&$Rmag_B$&$Rmag_{ref}$&Err Rmag&JD\\
\hline
18.308300 & 17.916424 & 19.959659 & 0.124381 & 050.69800\\
18.323450 & 17.937721 & 19.929005 & 0.034650 & 062.73200\\
18.307941 & 17.943148 & 19.933155 & 0.049381 & 068.69100\\
18.277621 & 17.921927 & --------- & 0.022007 & 074.71047\\
18.317622 & 17.978474 & 19.906741 & 0.046542 & 100.66100\\
18.340800 & 17.996533 & 19.931855 & 0.028322 & 103.73100\\
18.343249 & 18.001368 & 19.906263 & 0.018813 & 105.76800\\
18.342268 & 18.016725 & 20.028774 & 0.028204 & 109.72400\\
18.353127 & 18.013164 & 19.943019 & 0.147739 & 111.65300\\
18.346979 & 17.990860 & 19.911710 & 0.010532 & 123.78600\\
18.339169 & 17.959495 & --------- & 0.034687 & 134.75378\\
18.387632 & 18.042124 & 19.966310 & 0.011901 & 138.58600\\
18.385108 & 18.037049 & 19.931430 & 0.012798 & 142.76500\\
18.415253 & 18.066928 & 19.936043 & 0.018701 & 155.43300\\
18.350806 & 17.993212 & --------- & 0.027625 & 156.58644\\
18.363621 & 18.000546 & --------- & 0.020099 & 158.59506\\
18.360164 & 17.996483 & --------- & 0.010037 & 160.60051\\
18.409971 & 18.072709 & 19.948912 & 0.013566 & 176.41200\\
18.394950 & 18.070155 & 19.918187 & 0.010659 & 182.41100\\
18.435319 & 18.079750 & --------- & 0.037845 & 215.53896\\
18.421549 & 18.055671 & 19.896153 & 0.010842 & 219.38300\\
18.417204 & 18.075555 & --------- & 0.037845 & 221.36005\\
18.460061 & 18.094696 & 19.930324 & 0.038369 & 227.41000\\
18.459076 & 18.113779 & 19.938801 & 0.093712 & 228.37800\\
18.405639 & 18.072108 & --------- & 0.063487 & 235.38689\\
18.409954 & 18.081878 & --------- & 0.018256 & 239.39029\\
18.459080 & 18.126627 & 19.937951 & 0.010008 & 247.42200\\
18.440717 & 18.107298 & 19.904796 & 0.010067 & 254.39000\\
18.429906 & 18.072435 & --------- & 0.091177 & 260.41507\\
18.487492 & 18.093200 & --------- & 0.066955 & 276.40500\\
\hline
\end{tabular}

\end{table*}
\end{center}

\scriptsize
\begin{center}
\begin{table*}
{\bf Table.3.} Photometry of  two images of SDSS J1206+4332 quasar and a reference star.
\centering
\begin{tabular}{|c|c|c|c|c|}
\hline
$Rmag_A$&$Rmag_B$&$Rmag_{ref}$&Err Rmag&JD\\
\hline
19.509589 & 19.096939 & 19.562536 & 0.062694 & 006.72319\\
19.458226 & 19.031157 & 19.575507 & 0.075527 & 012.73698\\
19.480679 & 19.029276 & 19.539772 & 0.040385 & 016.66442\\
19.481074 & 19.031575 & 19.501510 & 0.010037 & 018.69594\\
19.498675 & 19.030665 & 19.572343 & 0.072392 & 048.69698\\
19.445953 & 18.967233 & 19.497185 & 0.010581 & 050.66706\\
19.452695 & 18.985873 & 19.519436 & 0.021286 & 053.73665\\
19.490599 & 19.001164 & 19.453006 & 0.048676 & 055.77248\\
19.449798 & 19.013616 & 19.526555 & 0.027773 & 061.65876\\
19.438714 & 19.048370 & 19.529403 & 0.030448 & 073.79091\\
19.563541 & 19.177637 & 19.512119 & 0.015220 & 088.59046\\
19.490950 & 19.112350 & 19.516885 & 0.019072 & 092.76935\\
19.345731 & 19.145619 & 19.454153 & 0.047554 & 126.41778\\
19.318989 & 19.138562 & 19.532942 & 0.033810 & 132.41905\\
19.265193 & 19.097541 & 19.597853 & 0.097722 & 169.38858\\
19.346323 & 19.165073 & 19.525828 & 0.027097 & 177.49348\\
19.341391 & 19.152732 & 19.587948 & 0.087875 & 178.38291\\
19.444082 & 19.135932 & 19.522717 & 0.024232 & 204.39591\\
19.456705 & 19.139128 & 19.494133 & 0.011932 & 206.47791\\
19.476497 & 19.165282 & 19.520598 & 0.022319 & 220.46142\\
19.458248 & 19.152088 & 19.467670 & 0.034457 & 224.38148\\
19.463009 & 19.133444 & 19.528405 & 0.029507 & 226.41176\\
19.489138 & 19.153896 & 19.544838 & 0.045311 & 230.41777\\
19.499540 & 19.151204 & 19.377492 & 0.123557 & 232.41145\\
19.475140 & 19.120366 & 19.341953 & 0.159005 & 234.46066\\
19.490768 & 19.109479 & 19.534108 & 0.034926 & 240.41984\\
19.506084 & 19.096243 & 19.498606 & 0.010205 & 244.40898\\
19.500106 & 19.028428 & 19.390230 & 0.110865 & 265.40616\\
19.449340 & 18.985593 & 19.512254 & 0.015322 & 269.39189\\
19.502726 & 19.039390 & 19.473984 & 0.028473 & 281.37866\\
19.507411 & 19.054861 & 19.307993 & 0.192910 & 283.37888\\
\hline
\end{tabular}
\end{table*}
\end{center}

\scriptsize
\begin{center}
\begin{table*}
{\bf Table.3.} Photometry of  two images of SDSS J1335+1138 quasar and a reference star.
\centering
\begin{tabular}{|c|c|c|c|c|}
\hline
$Rmag_A$&$Rmag_B$&$Rmag_{ref}$&Err Rmag&JD\\
\hline
18.103033 & 19.366769 & --------- & 0.020000 & 30.724810\\
18.173849 & 19.445618 & --------- & 0.020000 & 35.773720\\
18.168048 & 19.507642 & 19.811958 & 0.017249 & 39.739070\\
18.138333 & 19.480296 & 19.827976 & 0.010191 & 42.774090\\
18.158132 & 19.434327 & --------- & 0.020000 & 43.732740\\
18.116399 & 19.338482 & --------- & 0.020000 & 49.774310\\
18.128543 & 19.508473 & 19.835769 & 0.013971 & 55.622420\\
18.132991 & 19.461828 & --------- & 0.020000 & 61.534550\\
18.094535 & 19.430734 & --------- & 0.020000 & 67.730140\\
18.109781 & 19.461155 & 19.816665 & 0.013688 & 100.67228\\
18.105817 & 19.439126 & --------- & 0.020000 & 106.51100\\
18.109180 & 19.381934 & --------- & 0.020000 & 110.62587\\
18.098207 & 19.397216 & --------- & 0.020000 & 114.58660\\
18.110809 & 19.398308 & --------- & 0.020000 & 116.54903\\
18.137283 & 19.435018 & --------- & 0.020000 & 120.53586\\
18.099405 & 19.501067 & 19.810479 & 0.018473 & 126.52229\\
18.103190 & 19.397234 & --------- & 0.020000 & 134.55204\\
18.086870 & 19.403923 & --------- & 0.020000 & 136.39267\\
18.101415 & 19.440893 & --------- & 0.020000 & 138.40052\\
18.094541 & 19.434533 & --------- & 0.020000 & 140.38453\\
18.095505 & 19.473765 & 19.828225 & 0.010241 & 155.41550\\
18.101533 & 19.475559 & 19.835056 & 0.013482 & 157.45223\\
18.076486 & 19.352538 & --------- & 0.020000 & 164.41317\\
18.082410 & 19.449831 & 19.824963 & 0.010035 & 179.41976\\
18.086675 & 19.448914 & 19.826855 & 0.017975 & 181.40853\\
18.088464 & 19.443357 & 19.840949 & 0.011272 & 183.41002\\
18.063292 & 19.420842 & 19.820810 & 0.015805 & 199.42655\\
18.017446 & 19.390862 & 19.838252 & 0.080799 & 213.41659\\
18.052497 & 19.417072 & 19.745833 & 0.012022 & 214.39304\\
18.045362 & 19.375354 & 19.832686 & 0.020918 & 228.41725\\
18.069432 & 19.327107 & 19.844385 & 0.016030 & 232.38603\\
18.057617 & 19.398826 & 19.838541 & 0.038127 & 234.37637\\
\hline
\end{tabular}
\end{table*}
\end{center}

\scriptsize
\begin{center}
\begin{table*}
{\bf Table.3.} Photometry of  two images of SDSS J1353+0118 quasar and a reference star.
\centering
\begin{tabular}{|c|c|c|c|c|}
\hline
$Rmag_A$&$Rmag_B$&$Rmag_{ref}$&Err Rmag&JD\\
\hline
17.058143 & 18.078725 & --------- & 0.062755 & 035.76942\\
17.027750 & 18.105148 & 18.374450 & 0.035975 & 039.74420\\
17.004627 & 18.077990 & 18.428243 & 0.021679 & 042.77780\\
16.968521 & 18.023901 & --------- & 0.069111 & 043.72839\\
17.036874 & 18.072364 & --------- & 0.058642 & 049.77080\\
17.036695 & 18.077008 & 18.429024 & 0.022376 & 053.68090\\
17.016458 & 18.085246 & 18.395050 & 0.017169 & 055.62640\\
16.990376 & 18.025689 & --------- & 0.057302 & 059.57378\\
17.058655 & 18.088771 & --------- & 0.048000 & 061.52997\\
16.964883 & 18.017243 & --------- & 0.040402 & 067.73733\\
16.981154 & 18.017522 & 18.415777 & 0.012076 & 100.67650\\
17.109052 & 18.194647 & --------- & 0.164170 & 106.51640\\
16.993320 & 18.077777 & --------- & 0.028363 & 110.62981\\
17.002714 & 18.107406 & --------- & 0.024942 & 114.59100\\
17.026681 & 18.117991 & --------- & 0.054771 & 116.55300\\
17.002861 & 18.087047 & --------- & 0.023093 & 120.53156\\
16.976714 & 18.029956 & --------- & 0.020771 & 122.53589\\
16.970766 & 18.151632 & 18.414655 & 0.011484 & 126.52910\\
16.990348 & 18.087793 & --------- & 0.027337 & 140.38854\\
16.968081 & 18.051361 & 18.336053 & 0.073636 & 155.42000\\
16.970249 & 18.026965 & 18.384892 & 0.026106 & 173.44620\\
16.966204 & 18.063215 & 18.344583 & 0.065195 & 175.39800\\
16.958724 & 18.024720 & 18.400829 & 0.012918 & 179.41520\\
16.975812 & 18.050705 & 18.385252 & 0.025774 & 181.40410\\
16.975648 & 18.037233 & 18.366688 & 0.043484 & 183.40570\\
17.009698 & 18.109882 & --------- & 0.028265 & 185.39787\\
16.974296 & 18.015906 & 18.404894 & 0.010812 & 189.42540\\
16.953534 & 18.011738 & 18.433626 & 0.026571 & 199.43070\\
16.921712 & 18.002514 & --------- & 0.099206 & 201.43826\\
16.950484 & 17.994469 & 18.491207 & 0.082805 & 213.42560\\
16.969916 & 18.041750 & 18.442072 & 0.034543 & 214.38850\\
16.963528 & 18.037533 & 18.444302 & 0.036684 & 216.42800\\
16.976914 & 18.056241 & 18.388224 & 0.023063 & 220.41840\\
16.946127 & 18.035193 & 18.394833 & 0.017347 & 228.42160\\
16.960770 & 18.070003 & 18.418786 & 0.013986 & 232.38040\\
16.961527 & 18.025517 & 18.410923 & 0.010181 & 234.37150\\
16.950683 & 18.015915 & 18.424231 & 0.018214 & 238.35420\\
16.953166 & 18.031116 & --------- & 0.070280 & 240.37300\\
\hline
\end{tabular}
\end{table*}
\end{center}

\end{document}